\documentclass[iicol,sn-mathphys]{sn-jnl}
\usepackage{bm,graphicx}

\newcommand{\beq}{\begin{equation}}
\newcommand{\eeq}{\end{equation}}
\newcommand{\beqa}{\begin{eqnarray}}
\newcommand{\eeqa}{\end{eqnarray}}

\newcommand{\abs}[1]{\vert#1\vert}
\newcommand{\comport}[2]{\mathrel{\mathop{#1}\limits_{#2}}}
\renewcommand{\e}{{\rm e}}
\newcommand{\eff}{{\rm eff}}
\newcommand{\eps}{\varepsilon}
\newcommand{\half}{{\scriptstyle{\frac{\scriptstyle1}{\scriptstyle2}}}}
\newcommand{\mean}[1]{\langle#1\rangle}
\newcommand{\prob}[1]{{\rm Prob}(#1)}
\newcommand{\s}{\sigma}
\newcommand{\sign}{\mathop{\rm sign}}
\newcommand{\typ}{{\rm typ}}
\renewcommand{\w}[1]{{\tilde#1}}
\renewcommand{\z}{\zeta}
\renewcommand{\H}{{\cal H}}
\newcommand{\I}{{\rm I}}
\newcommand{\N}{{\cal N}}

\begin{document}

\title{Speech perception: a model of word recognition}

\author*[1]{\fnm{Jean-Marc} \sur{Luck}}\email{jean-marc.luck@ipht.fr}
\author[2]{\fnm{Anita} \sur{Mehta}}\email{anita.mehta@oxon.org}

\affil[1]{\orgname{Universit\'e Paris-Saclay, CNRS \& CEA},
\orgdiv{Institut de Physique Th\'eorique},
\city{91191~Gif-sur-Yvette}, \country{France}}

\affil[2]{\orgdiv{Faculty of Linguistics, Philology and Phonetics},
\orgname{Clarendon Institute, Walton Street},
\city{Oxford OX1 2HG}, \country{United Kingdom}}

\abstract{We present a model of speech perception which takes into account effects of correlations between sounds.
Words in this model correspond to the attractors of a suitably chosen descent dynamics.
The resulting lexicon is rich in short words, and much less so in longer ones,
as befits a reasonable word length distribution.
We separately examine the decryption of short and long words in the presence of mishearings.
In the regime of short words, the algorithm either quickly retrieves a word, or proposes another valid word.
In the regime of longer words, the behaviour is markedly different.
While the successful decryption of words continues to be relatively fast,
there is a finite probability of getting lost permanently,
as the algorithm wanders round the landscape of suitable words without ever settling on one.}

\maketitle

\section{Introduction}
\label{intro}

The perception of speech is of immense importance in linguistics
(see~\cite{grzybek,hay,lad,G+J} for recent overviews of the fields of phonetics and phonology).
The main quantitative subfield is that of computational linguistics.
Our approach is rather different.
We use the tools of statistical physics to make predictions on observables of interest
and explore the effects of mishearings on the process of decrypting words that have been heard.
The goal is thus not the {\it exact} modelling of the decryption of a given word, but an
exploration of some of the statistical characteristics of how we hear and interpret speech.
In this sense, our approach might
be construed as the polar opposite to that of computational linguistics.

In earlier work~\cite{us}, we had presented a pheno\-menological model of word recognition.
This was based on the approximation that every sound was equally probable,
and that each sound could be freely juxtaposed with any other, in word formation.
Despite this over-simplification,
it turned out that one could make several reasonable predictions on the basis of counting (or `phase-space') arguments,
constructed by using realistic word length distributions obtained via~\cite{leipzig}.
These included anticipation thresholds
(prefix lengths after which one could correctly predict the corresponding words),
as well as some simple effects of mishearings which might modify the above.
Where these arguments were not successful at more than a descriptive level was in the modelling of
dynamical effects, to do with, say, the slowing down anticipated for the retrieval of longer words
as well as the phenomenon of `getting lost', where one is simply unable to retrieve a badly misheard word.
As will be seen, the current model has enough complexity to ensure that these goals are achieved.

We first recall that the smallest distinctive unit of speech is called a phoneme~\cite{grzybek,hay,lad,G+J}.
These phonemes, divided into vowels and consonants, are specific to a given language.
Their total number ranges between a minimum
of 11 and a maximum of 160 across world languages~\cite{hay}.
Typically, languages have 20 to 50 phonemes; English, for example, has 44.
Lahiri and co-workers~\cite{lahiri2} sought to make this deconstruction more universal,
by relating perceived sounds to anatomical {\it features}, so that
language-independent frameworks for analysis could be set up.
The universality of this scheme provided the inspiration for the present model,
which is also set up in terms of a language-independent formalism; however, our model is far from
being able to incorporate the subtle details of features.
We refer instead to the elementary units of speech simply as `sounds' which are
meant to be universal across languages,
along the lines suggested by Lahiri et~al.~\cite{lahiri1}.

A quantity central to our study is the distribution of word lengths
(see~\cite[Chapter~2]{grzybek} for a comprehensive review).
The overwhelming majority of available data concerns the length distribution
of unique written words, where each word of the lexicon is counted once,
and its length is defined as the number of its letters.
There are typically 100,000 words in the lexicons of most languages.
It appears that word length distributions in most active languages are
characterised by two main features:
a very rapid initial increase and a more or less steep decay with length.
In earlier work~\cite{us}, we put forward the use of the Gamma distribution
\beq
p_n=C\,n^\alpha\,\e^{-\beta n}
\label{gamma}
\eeq
as an efficient way to describe these features.
This distribution has a minimal number of parameters,~$\alpha$ and $\beta$,
describing respectively the rise and fall of the distribution
(the prefactor $C$ is determined by normalisation).
We showed~\cite{us} that this distribution provided a very good
parametrisation for European as well as Asian languages
which had between them a rather broad range of average word lengths.
We therefore argued that the Gamma distribution~(\ref{gamma}) has a universal
validity across most world languages.

In the context of the present model, the relevant concept of a word length is the number of its sounds.
Fortunately, it turns out that there is a high degree of correlation between the length distributions
of written words (numbers of letters, denoted by $n$) and spoken words (numbers of sounds,
i.e., phonemes, denoted by $m$), for most languages~\cite{hay,lad}.
As a matter of fact, sounds and letters enjoy a near one-to-one correspondence
in ancient languages like Latin or Greek,
so that their length distributions for written and spoken words are virtually identical.
This is referred to as `phonemic orthography'.
We therefore expect that this high degree of correspondence between
phonology and orthography would persist in modern Western languages.
This expectation has been confirmed~\cite{us1} via
an analysis of this question in the case of three languages (German, Dutch and English)
which are well-documented in the CELEX database~\cite{celex}.
Our main conclusion is that there is a fixed ratio $r$ close to unity
between the numbers of sounds and letters in typical words:
\beq
m\approx rn\qquad(r\approx0.85).
\label{mrn}
\eeq
We obtain nearly identical values of $r$ for unique words (when every word of the lexicon is counted once)
and token words (when words are counted according to their frequencies of occurrence).
Furthermore, the above value of the ratio $r$ is common to the three languages we have tested,
and thus appears to have a high degree of universality.

Sounds are therefore the essential building blocks of the present model; unlike the case of our
previous work, {\it these sounds are correlated, so that only compatible sounds can be neighbours}.
This marks a major advance from the earlier picture~\cite{us} of a flat landscape of sounds, where
the juxtaposition of sounds next to each other could be done independently of their nature.
In the present model, sounds are represented by strings of six spins, whose interactions are
described by two terms, comprising
a disordered Hamiltonian one (which will be seen to be adequate for short words),
as well as a variable-range non-Hamiltonian one (which will be needed for longer words).
Under suitable stochastic descent dynamics,
the attractors of the spin system so defined are the words of the lexicon.

The dynamics of word retrieval is the central concern of this paper.
It corresponds to the decryption of a misheard word via the above descent dynamics.
Words are thus identified as the fixed points of the dynamics.
This algorithmic viewpoint on the task of decrypting a mispronounced or misheard word is the
second main advance of this work with respect to its predecessor~\cite{us}.

The setup of this paper is as follows.
Section~\ref{typical} is devoted to typical words,
whose length $m$ is close to the mean word length $\mean{m}$ of the language under scrutiny.
In Section~\ref{typmodel} we describe our approach,
consisting of modelling words as strings of Ising spins,
endowing the spin chain so obtained with a Hamiltonian involving short-range quenched disorder,
and the lexicon as the set of all attractors of the corresponding descent dynamics.
In Section~\ref{typdecrypt} we describe how to model the decryption of mispronounced or misheard words
in this setting, and investigate the main characteristics of the decryption stage.
Section~\ref{long} is devoted to long words,
whose length $m$ is much larger than the mean word length $\mean{m}$.
These words become increasingly rare as a function of their length.
In Section~\ref{longmodel} we describe how this rarity is modelled; a variable-range non-Hamiltonian contribution,
with coupling constant $g$ and interaction range $\ell$, is superposed with the short-range
Hamiltonian, which results in the breaking of reciprocity and detailed balance.
A sharp crossover results from the competition between these two terms.
The ensuing `phase diagram' in the $\ell$--$g$ plane is investigated in Section~\ref{longphase}.
The explicit correspondence between the parameters of the present model
and realistic word length distributions is also given.
The section concludes with a presentation of the main properties of the decryption of misheard long words
(Section~\ref{longdecrypt}).
We discuss our findings in Section~\ref{disc},
while technical details of some facets of our model are presented in four appendices.

\section{Typical words}
\label{typical}

In this section we consider typical spoken words,
whose length (number of sounds) $m$ is not too different from the mean word length $\mean{m}$
of the language under consideration.

\subsection{Modelling the lexicon}
\label{typmodel}

As recalled in Section~\ref{intro}, languages typically have 20 to 50 different phonemes.
In this work we use Ising spins, i.e., binary variables $\s_i=\pm1$, as our basic degrees of freedom.
A string of 6 spins is sufficient to encode all phonemes of such typical languages (since $2^6=64$).
From now on, we choose this correspondence,
and represent a spoken word consisting of $m$ sounds by an array~of
\beq
L=6m
\label{lm}
\eeq
spins.
There are typically 100,000 unique words in the lexicons of most languages,
so that more than 10,000 words have the most probable word length.
We must therefore devise a model whose dynamics admit such a large number of fixed points.
It turns out that the physics of one-dimensional disordered systems provides such examples~\cite{dg,msj}.

We accordingly consider the Hamiltonian
\beq
\H=-\sum_{i=1}^L(J_i\s_i\s_{i+1}+h_i\s_i)
\label{ham}
\eeq
on an open chain of length $L$ with free boundary conditions.
The Hamiltonian~(\ref{ham}) has been chosen from among a range of
disordered and frustrated one-dimensional models because of its minimality.
The exchange couplings $J_i$ between nearest neighbours, and the magnetic fields $h_i$, are
independent quenched random variables.
From our point of view, the random exchanges $J_i$ model the complex couplings between neighbouring
features -- for example, some may be forbidden while others may be favoured~\cite{lahiri1}.
The local fields $h_i$ embody further constraints,
in the sense of forcing the sounds into a shape resembling a real word.

In order to have as unbiased a lexicon as possible,
we prevent the occurrence of any kind of order,
by postulating that the distributions $\rho_J$ and $\rho_h$ are even,
i.e., symmetric under $J_i\to -J_i$ or $h_i\to -h_i$.
In this situation, the ensemble of random Hamiltonians of the form~(\ref{ham})
is invariant under any gauge transformation~\cite{toulouse},
i.e., any change of variables of the form
\beq
\w\s_i=\eps_i\s_i,\quad
\w h_i=\eps_i h_i,\quad
\w J_i=\eps_i\eps_{i+1}J_i,
\label{gauge}
\eeq
where $\eps_i=\pm1$ are arbitrary signs attached to the sites.
For definiteness, the $J_i$ and the $h_i$ are drawn from the symmetric uniform distributions
\beq
\left\{
\begin{matrix}
\rho_J(J_i)=\frac{1}{2}\hfill & (-1\le J_i\le+1),\hfill\cr
\rho_h(h_i)=\frac{1}{2w}\quad\hfill & (-w\le h_i\le+w).\hfill
\end{matrix}
\right.
\label{rhos}
\eeq
throughout this paper.
The width $w$ of the random field distribution is kept as a free parameter.

The dynamics modelling word decryption are chosen to be
zero-temperature single-spin-flip dynamics, associated with the random Hamiltonian $\H$.
Each spin is successively updated according to the dynamical rule
\beq
\s_i\to\sign\eta_i,
\label{dy}
\eeq
where the total field $\eta_i$ acting upon spin $\s_i$ reads
\beq
\eta_i=-\frac{\partial\H}{\partial\s_i}=J_{i-1}\s_{i-1}+J_i\s_{i+1}+h_i.
\label{eta}
\eeq
Two different kinds of sequential single-spin updates will be considered : random updates,
where at each time step $\delta t=1/L$, the spin to be updated is chosen uniformly at random over the system,
and ordered updates, where spins $\s_1,\dots,\s_L$ are successively updated at each sweep.

Equation~(\ref{dy}) corresponds to descent dynamics,
where each spin flip changes the total energy by a negative amount,
\beq
\delta\H=-2\abs{\eta_i}.
\eeq
All attractors are therefore fixed points,
obeying
\beq
\s_i=\sign\eta_i,\quad\hbox{i.e.,}\quad\s_i\eta_i>0
\label{fp}
\eeq
for all spins $(i=1,\dots,L)$.
These attractors, where each spin is aligned along the total field acting upon it,
are also known as one-spin-flip metastable configurations~\cite{dg,msj,ns}.
In the context of our model, they are precisely the words of the lexicon.
That is, for a given draw of the random Hamiltonian $\H$,
the corresponding lexicon is the set of all attractors obeying~(\ref{fp}).
The size of this lexicon, i.e., the number of attractors,
typically grows exponentially with the system length, as
\beq
\N\sim\exp(SL),
\label{nexp}
\eeq
where $S$ is the configurational entropy~\cite{gibbsd,jackle,palmer}
(see~\cite{DS} for a review).
A more accurate statement of the above growth law relies on the definition of the typical number of attractors,
\beq
\N_\typ=\exp\mean{\ln\N}.
\label{ntypdef}
\eeq
The growth law~(\ref{nexp}) then becomes a usual law of extensivity of the form
\beq
\mean{\ln\N}\approx SL.
\eeq

\begin{figure}[!ht]
\begin{center}
\includegraphics[angle=0,width=0.9\linewidth,clip=true]{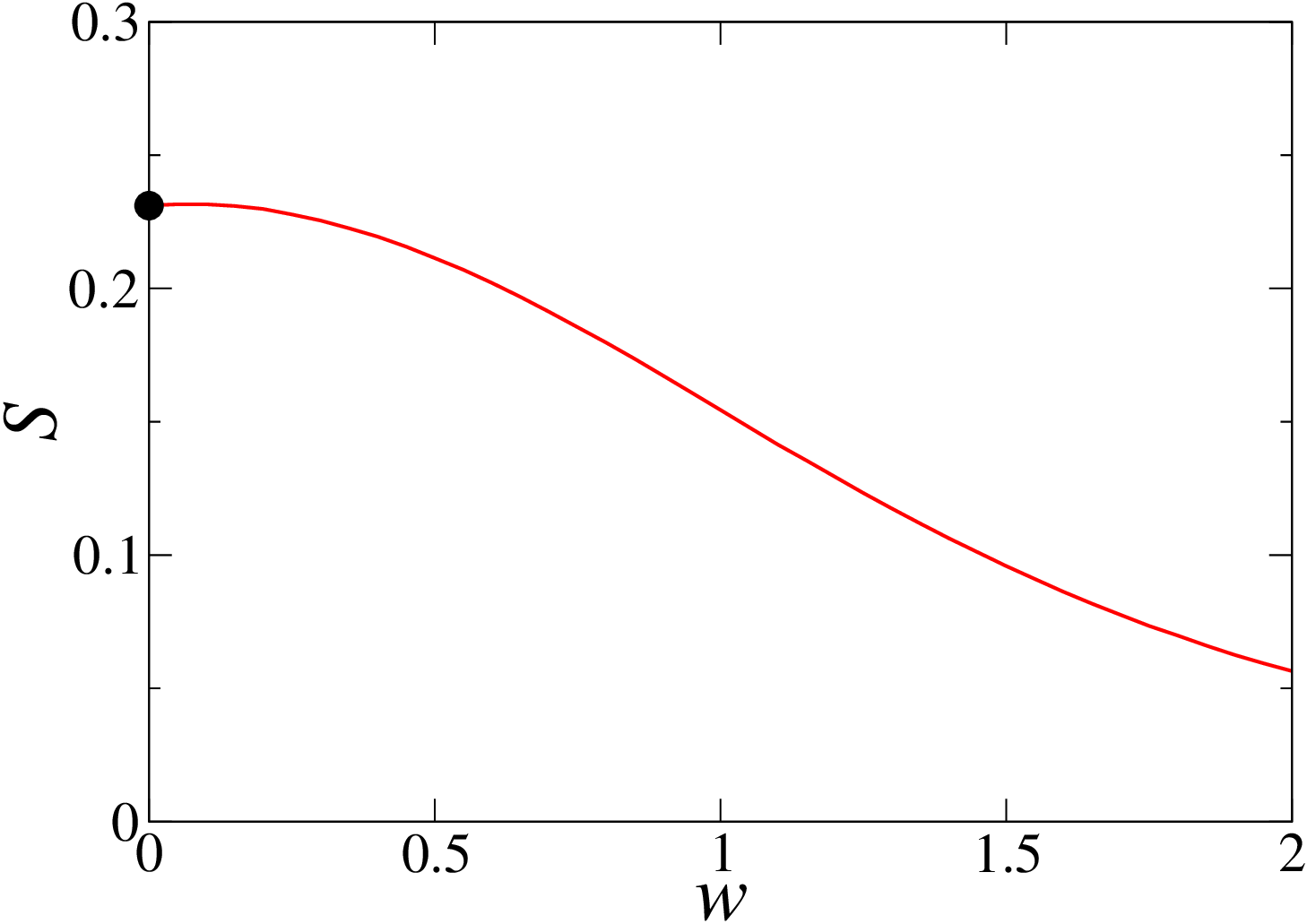}
\caption{\small
Configurational entropy $S$ against width $w$ of the random field distribution.
Symbol: exactly known value $S_0$ in the absence of random fields (see~(\ref{szero})).}
\label{sent}
\end{center}
\end{figure}

In the absence of random fields $(w=0)$,
the model boils down to the spin-glass chain (see Appendix~\ref{appsg}).
The latter model is not frustrated.
Its attractors have been studied by Derrida and Gardner~\cite{dg}.
The corresponding entropy reads $S_0=(\ln 2)/3\approx0.231049$ (see~(\ref{szero})).
In the presence of random fields, however, $S$ is not known exactly.
We have devised a recursive computer algorithm, freely inspired from~\cite{msj},
which enumerates all attractors for a given draw of the Hamiltonian $\H$ of a system of length $L$.
Figure~\ref{sent} shows a plot of the configurational entropy $S$ thus obtained against $w$.
The asymptotic value of $S$ is essentially reached at $L=30$.
The configurational entropy decreases continuously from $S_0$ at $w=0$ (represented by the symbol in the figure)
and slowly goes to zero at large $w$.
In the limit of an infinitely large $w$, the model is no longer frustrated,
since each spin $\s_i$ is aligned with its random field $h_i$.

It is of interest to compare the configurational entropy plotted in Figure~\ref{sent} with actual linguistic data.
Let us consider for definiteness English and French (see the first two lines of Table~\ref{abtab}).
The mean lengths of spoken words
are respectively $\mean{m}_{\rm E}\approx7.1$ (i.e., $\mean{L}_{\rm E}=6\mean{m}\approx43$) for English
and $\mean{m}_{\rm F}\approx8.6$ (i.e., $\mean{L}_{\rm F}=6\mean{m}\approx52$) for French.
Assuming for simplicity that about 10,000 unique words have these typical lengths,
the corresponding configurational entropies can be estimated as $S_{\rm E}\approx(\ln
10,000)/43\approx0.22$ and $S_{\rm F}\approx(\ln 10,000)/52\approx0.18$.
The agreement between these estimates and values of the configurational entropy shown in
Figure~\ref{sent} for moderate values of~$w$,
i.e., $w_{\rm E}\approx0.4$ and $w_{\rm F}\approx0.8$,
provides strong evidence of the adequacy of our scheme for modelling realistic lexicons.

\subsection{Decrypting misheard words}
\label{typdecrypt}

Within the present setting,
the task of decrypting a mispronounced or misheard word is modelled as follows:
\beq
W\comport{\xrightarrow{\hspace{30pt}}}{\rm alteration}\Sigma
\comport{\xrightarrow{\hspace{30pt}}}{\rm decryption}X.
\label{scheme}
\eeq

\begin{itemize}

\item
{\it Alteration stage.}
Assume that the speaker intends to pronounce a word $W=\{\s_i^W\}$ of the lexicon
corresponding to a fixed draw of $\H$.
The mispronounced or misheard signal $\Sigma=\{\s_i^\Sigma\}$ perceived by the listener
is modelled by altering $W$ without changing its length $L$.
Unless otherwise stated,
$\Sigma$ is derived from $W$ by independently flipping each of its $L$ spins with a fixed
probability $p$, referred to as the density of mishearings:
\beq
\s_i^\Sigma=\left\{
\begin{matrix}
\hfill\s_i^W\quad &\hbox{with prob.~$1-p$},\hfill\cr
\hfill-\s_i^W\quad &\hbox{with prob.~$p$}.\hfill
\end{matrix}
\right.
\label{sigmap}
\eeq

\item
{\it Decryption stage.}
The listener applies the decrypting task modelled by the descent dynamics~(\ref{dy})
corresponding to the same draw of $\H$,
to the perceived signal $\Sigma$, with either a random or an ordered spin update.
This scheme yields a reconstructed word $X=\{\s_i^X\}$ whose length $L$ is that of $W$,
and which is valid in the sense that it belongs to the lexicon,
but may or may not coincide with the intended word $W$.

\end{itemize}

The alteration stage introduces a random number $N_0$ of initial mishearings,
equal to the distance between $W$ and $\Sigma$:
\beq
N_0=d(W,\Sigma)=\frac12\sum_{i=1}^L\left(1-\s_i^W\s_i^\Sigma\right).
\eeq
In the case of a uniform density of mishearings (see~(\ref{sigmap})),
$N_0$ may take any value between 0 and~$L$, and obeys the binomial law
\beq
\prob{N_0=k}={L\choose k}p^k(1-p)^{L-k}.
\label{nbin}
\eeq
We have in particular
\beq
\mean{N_0}=p L.
\label{nzero}
\eeq
Similarly, we denote by $N$
the random number of residual mishearings after the decryption stage has been completed,
equal to the distance between $W$ and $X$:
\beq
N=d(W,X)=\frac12\sum_{i=1}^L\left(1-\s_i^W\s_i^X\right).
\label{ndef}
\eeq
The decryption task is successful if all mishearings present in $\Sigma$ are healed,
so that the words $W$ and~$X$ coincide, i.e., $N=0$.
This takes place with a probability of success
\beq
P=\prob{X=W}=\prob{N=0}.
\label{psucc}
\eeq

A peculiarity of the present model is that the presence of a single residual mishearing ($N=1$)
is prohibited by the form~(\ref{ham}) of the Hamiltonian.
In other words, if the decryption is not successful,
then $W$ and $X$ must differ in at least two places, i.e., $N\ge2$.
As a consequence, we have
\beq
\mean{N}\ge2(1-P).
\label{n2p}
\eeq

\begin{figure}[!ht]
\begin{center}
\includegraphics[angle=0,width=0.9\linewidth,clip=true]{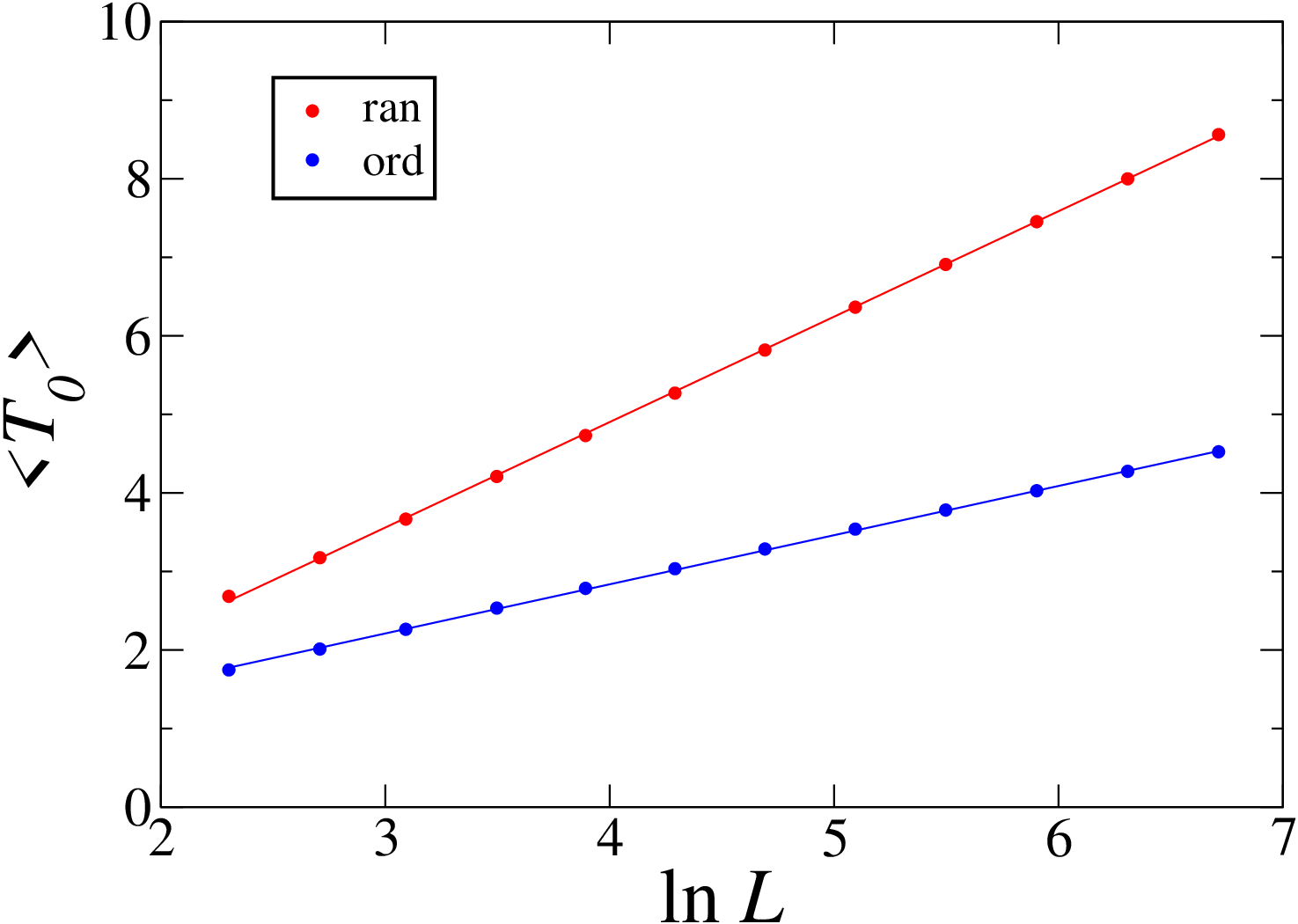}
\caption{\small
Mean time $\mean{T_0}$ taken by the descent dynamics~(\ref{dy})
to select a word $W$ from a disordered initial configuration,
plotted against $\ln L$ for $w=1$.
Red: Random update.
Blue: Ordered update.
Full lines: least-square fits with respective slopes 1.34 and 0.63.}
\label{starttime}
\end{center}
\end{figure}

We have performed numerical simulations of the decryption of misheard words modelled as above.
For a given draw of the Hamiltonian $\H$,
we first select a word $W$ by applying the descent dynamics~(\ref{dy}) to a disordered initial configuration,
and then model its decryption according to the scheme~(\ref{scheme}).
The dynamical process of selecting a word is fast (see Figure~\ref{starttime}).
The mean time it takes for the descent process to converge grows as
\beq
\mean{T_0}\approx A\ln L.
\label{tlog}
\eeq
This is the expected scaling for dynamics consisting of independent rearrangements of $L$ degrees of freedom,
where $T_0$ appears as the largest of~$L$ independent and exponentially distributed time variables.
The amplitude $A$ depends significantly on the updating scheme; the ordered update is twice as fast
($A\approx0.63$) as the random update ($A\approx1.34$).

The words $W$ selected by means of the above dynamical process
are not distributed uniformly over the lexicon.
In other words, the Edwards hypothesis, commonly used in the realm of granular and glassy materials
(see~\cite{edrev} and references therein),
which states that all attractors have equally sized basins of attraction under a complex enough dynamics,
is known to be weakly violated for one-dimensional systems such as the present one
(see~\cite{epjst} and references therein).
This small effect is immaterial for our purposes.

\begin{figure}[!ht]
\begin{center}
\includegraphics[angle=0,width=0.9\linewidth,clip=true]{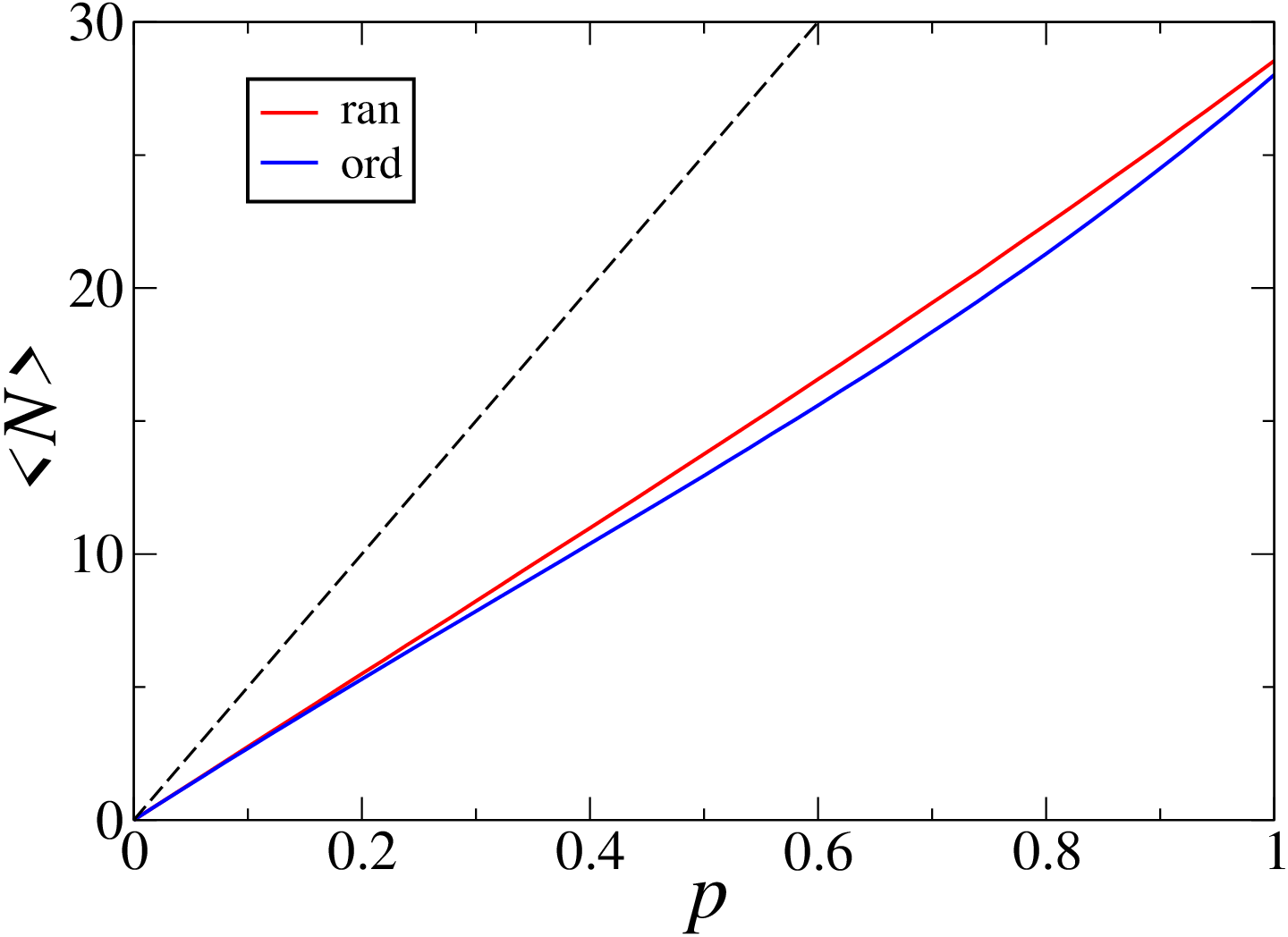}
\caption{\small
Mean number $\mean{N}$ of residual mishearings against density $p$ of initial mishearings, for
$L=50$ and $w=1$.
Red: Random update.
Blue: Ordered update.
Black dashed line: $\mean{N_0}$ (see~(\ref{nzero})).}
\label{ave}
\end{center}
\end{figure}

The mean number $\mean{N}$ of residual mishearings,
after the decryption stage has been completed,
is plotted in Figure~\ref{ave} against the density $p$ of initial mishearings
for $L=50$ and $w=1$, with random and ordered updates.
All data points obey the inequality
\beq
\mean{N}\le\mean{N_0},
\label{nineq}
\eeq
which says that the mean number of residual mishearings is less than the mean number of initial
mishearings in the heard word.
This is of course the least that one might expect from a decryption algorithm!
Besides this, $\mean{N}$ grows nearly linearly over the whole range of $p$,
and exhibits a weak dependence on the updating scheme.
As one might expect, the higher the number of initial mishearings, the more difficult it is to heal the system.

\begin{figure}[!ht]
\begin{center}
\includegraphics[angle=0,width=0.9\linewidth,clip=true]{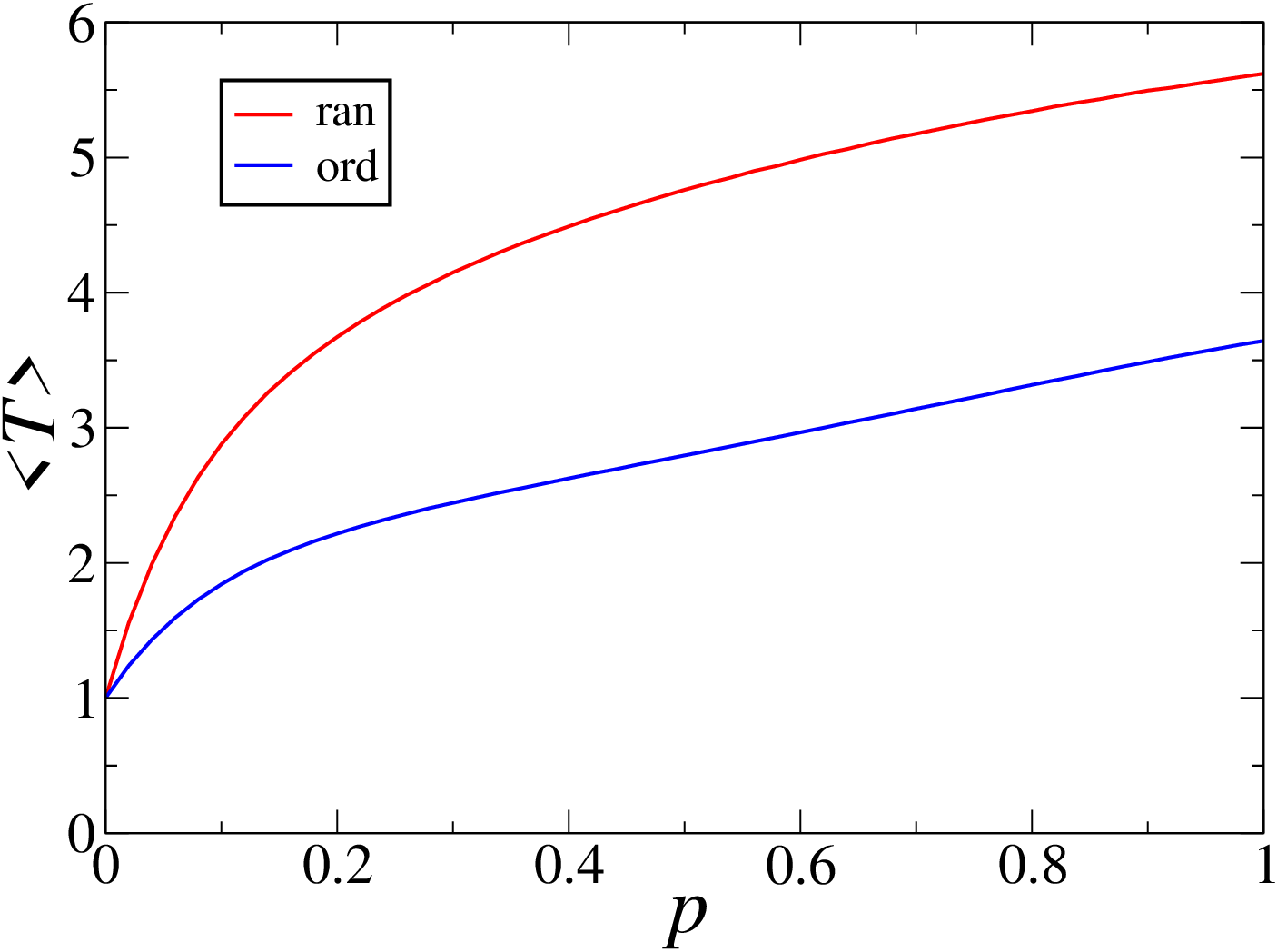}
\caption{\small
Mean running time $\mean{T}$ of the decryption process against density $p$ of mishearings.
Same parameters and conventions as in Figure~\ref{ave}.}
\label{time}
\end{center}
\end{figure}

The running time $T$ of the decryption process is comparable with the time $T_0$
taken by the descent dynamics to select a word from a random initial configuration.
In particular, both times grow logarithmically with the sample length~$L$.
For fixed $L$, Figure~\ref{time} shows that $\mean{T}$ increases steadily with~$p$, as might be expected.
Moreover, the ordered updating scheme is once again about twice as fast as the random one.
This, too, might be expected,
since the ordered clearing of mishearings as a word is `read' from left to right
hinders the random recurrence of mishearings in areas already cleared.

\begin{figure}[!ht]
\begin{center}
\includegraphics[angle=0,width=0.9\linewidth,clip=true]{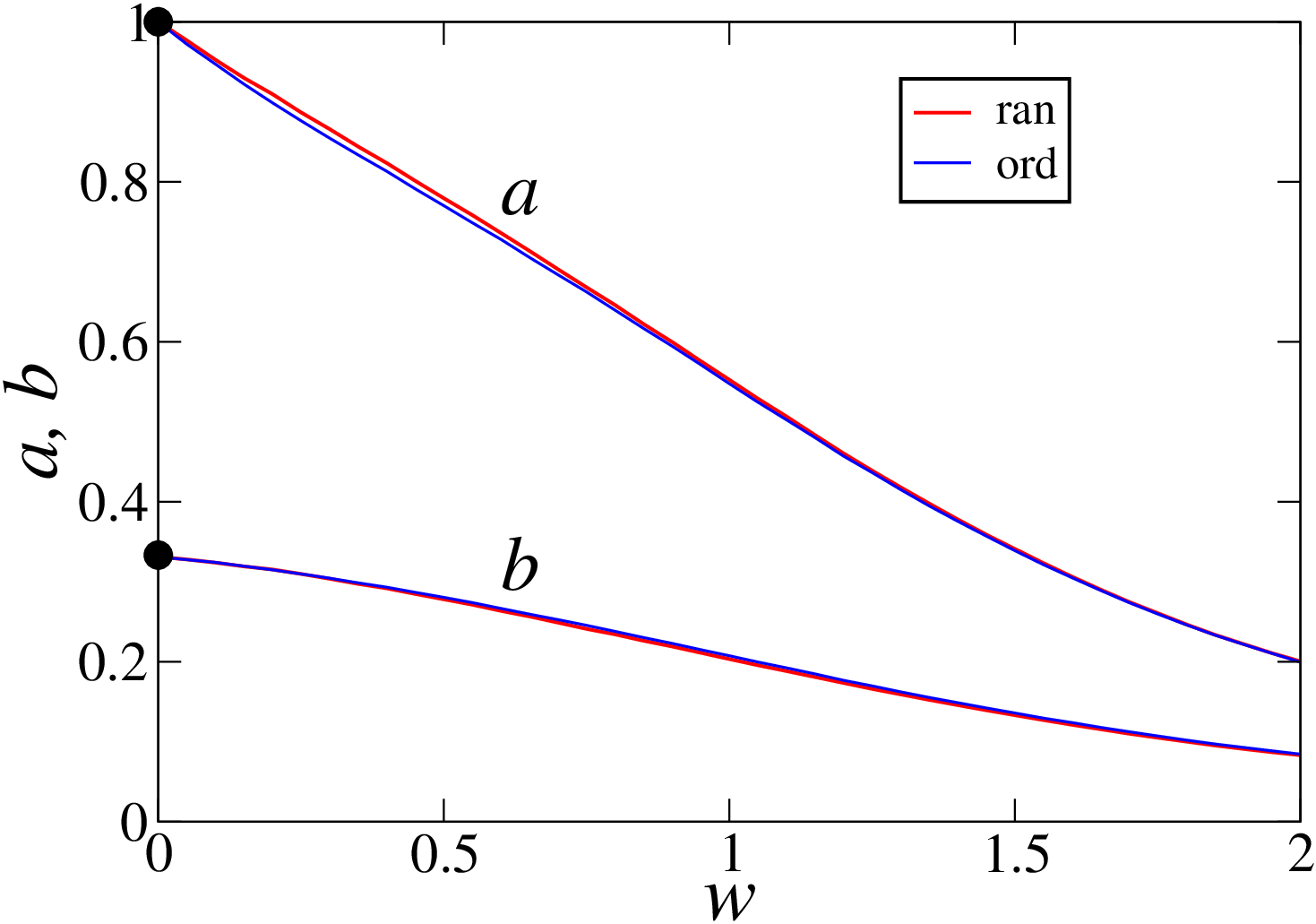}
\caption{\small
Amplitudes $a$ (upper curves) and $b$ (lower curves) entering the scaling results~(\ref{npsca}),
plotted against the width $w$ of the local field distribution.
Same parameters and conventions as in Figure~\ref{ave}.
Symbols: maximal values $a_0=1$ and $b_0=1/3$ in the absence of random fields.}
\label{abplot}
\end{center}
\end{figure}

We now focus on the scaling regime where the system length $L$ is large and the density~$p$ of
mishearings is small, so that their product $\mean{N_0}=p L$ is finite (see~(\ref{nzero})).
In this regime, the binomial law~(\ref{nbin}) becomes the Poisson law
\beq
\prob{N_0=k}\approx\e^{-p L}\,\frac{(p L)^k}{k!}.
\label{npoi}
\eeq
All over this regime,
the mean number of residual mishearings and the success probability scale as
\beq
\mean{N}\approx a p L,\quad
P\approx\exp(-b p L).
\label{npsca}
\eeq
The amplitudes $a$ and $b$ measure the efficiency of the decrypting algorithm
in the regime of dilute mishearings.
More precisely, if a single mishearing is introduced randomly into a valid word,
$b$ is the probability that this mishearing is not healed during the decryption stage,
while $a=\mean{N}$ is the mean distance between $W$ and $X$ (see~(\ref{ndef})).
In the case of a totally inefficient decryption,
$N=N_0$ is distributed according to the Poisson law~(\ref{npoi}),
and so we have $a=b=1$.
In the present setting, the inequalities~(\ref{n2p}) and~(\ref{nineq}) translate to
\beq
2b\le a\le 1.
\label{abineqs}
\eeq
Figure~\ref{abplot} shows a plot of the amplitudes $a$ and~$b$ against~$w$.
Both decrease monotonically from their maximal values $a_0=1$ and $b_0=1/3$
in the absence of random fields, that are derived in Appendix~\ref{appsg} (see~(\ref{abzero})).
The second inequality in~(\ref{abineqs}) saturates at $a_0=1$.
Increasing $w$, i.e., introducing more frustration into the model,
improves the efficiency of the decrypting algorithm.
Again, this is to be expected; the random field plays the role
of a pinning agent, and in a sense prevents the recurrence of random errors during updates.
The dependence of $a$ and $b$ on the updating scheme is even weaker than in Figure~\ref{ave},
with at most a relative difference of 2 percent.
Although much larger than statistical errors, this difference is hardly visible on the plot.

To end this section, we emphasise that the amplitudes $a$ and $b$ are significant for all parameter values;
our decryption algorithm is thus intrinsically imperfect, always leaving
behind a finite proportion of residual mishearings.
In the widely investigated field of error correction~\cite{hamming}
(see~\cite{KZ} for a modern overview of related matters), by contrast, an asymptotically perfect
decryption is achieved by adding a sufficient amount of
redundancy to the transmitted message.\footnote{The density of unresolved residual errors
may vanish as~$p^2$ or faster, depending on details of the error-correcting code used.}
This additional manipulation of the
transmitted string by adding redundancy is there justified, because the primary aim of the error
correction process is to get as near a perfect decryption as possible.

Our aims in this paper are rather different: by constructing a minimal model of word decryption,
which is amenable to simple analysis, we aim to illustrate the main static and dynamical mechanisms
involved in the process.
Improving the efficiency of our model by, for instance,
adding higher-order spin interactions would be counterproductive,
since it would make the analysis of our model somewhat opaque.

\section{Long words}
\label{long}

In this section we consider long words,
whose length (number of sounds) $m$ is significantly larger than the mean word length $\mean{m}$
of the language under consideration.

\subsection{Modelling long words}
\label{longmodel}

The model proposed in Section~\ref{typical} to describe typical words
has to be adapted in order to deal with the case of long words.
Increasing the system length $L$ alone is insufficient in several respects.
First, the configurational entropy has to be decreased significantly and in a tunable fashion,
in order to take into account the exponential decay~(\ref{gamma}) of the word length distribution.
Second, the decryption of long words is expected to give rise
to qualitatively novel dynamical features~\cite{us}.

We continue to identify words with the fixed points of some dynamics of the form~(\ref{dy}):
\beq
\s_i\to\sign\eta_i.
\label{dylong}
\eeq
In order to meet the requirements mentioned above, however,
we introduce a key new ingredient, that of choosing a total field $\eta_i$
which does {\it not} derive from a Hamiltonian; it therefore
breaks reciprocity and detailed balance.
The specific choice
\beqa
\eta_i&=&J_{i-1}\s_{i-1}+J_i\s_{i+1}+h_i
\nonumber\\
&+&g\sum_{j=i+1}^Lx^{j-i-1}\s_j
\label{etalong}
\eeqa
can be supported by the following phenomenological argument.
It is inspired by the effects of the constraints that limit the free evolution of the ends of a given
`prefix' (here we use this word in the sense of computer science rather than grammar -- a prefix is
a set of sounds comprising the earlier part of a word).
The possible endings of long words have to fit in a highly constrained way with their bulk.
In the context of our model, it is as if the
nature (`feature'~\cite{lahiri1}) of the $i$-th sound in a word of more than some typical length has
to satisfy both the prefix that precedes it, as well as the rest of the word that follows it.
The coupling constant~$g$ embodies the relative importance of the end of a long word compared to its prefix.
This picture is qualitatively similar to that described in
earlier work on the compacting dynamics of a column of grains~\cite{lm1,lm2}, where in the most
compacting (`jamming') limit, grains in the middle of a column are no longer free to choose their
orientations, but encounter frustration as they try to satisfy the orientations of grains below and above them.

The definition~(\ref{etalong}) of the total field~$\eta_i$ thus consists of two qualitatively
different components:

\begin{itemize}

\item
The first line coincides with~(\ref{eta}) and therefore derives from the random Hamiltonian~(\ref{ham}).
It is short-ranged, conservative, and involves quenched disorder.
As mentioned above, this component does not induce any kind of order whatever.

\item
The second line is totally directed and non-local, with a variable range; it is, however,
translationally invariant.
It breaks reciprocity (i.e., the principle of action and reaction) and detailed balance.
It also explicitly breaks the large gauge invariance of the random Hamiltonian~(\ref{ham}).
It turns out that this is a very effective way of reducing the configurational entropy.
Also, as written in~(\ref{etalong}),
the symmetry-breaking term favours long-range ferromagnetic order.
We emphasise, however, that
antiferromagnetic order (or any other type of prescribed order) could have been chosen as well,
because of the gauge invariance~(\ref{gauge}).

\end{itemize}

The symmetry-breaking component given in the second line of~(\ref{etalong}) involves two novel parameters,
the coupling constant $g$ and the interaction range $\ell$, such that
\beq
x=\exp(-1/\ell).
\label{xdef}
\eeq
In the $\ell\to0$ (i.e., $x\to0$) limit,
the expression~(\ref{etalong}) of the total field becomes local again,
as it involves only nearest-neighbour interactions:
\beq
\eta_i=J_{i-1}\s_{i-1}+(J_i+g)\s_{i+1}+h_i,
\label{etazero}
\eeq
while keeping its non-reciprocal character through the term $g\s_{i+1}$.

\begin{figure}[!ht]
\begin{center}
\includegraphics[angle=0,width=0.9\linewidth,clip=true]{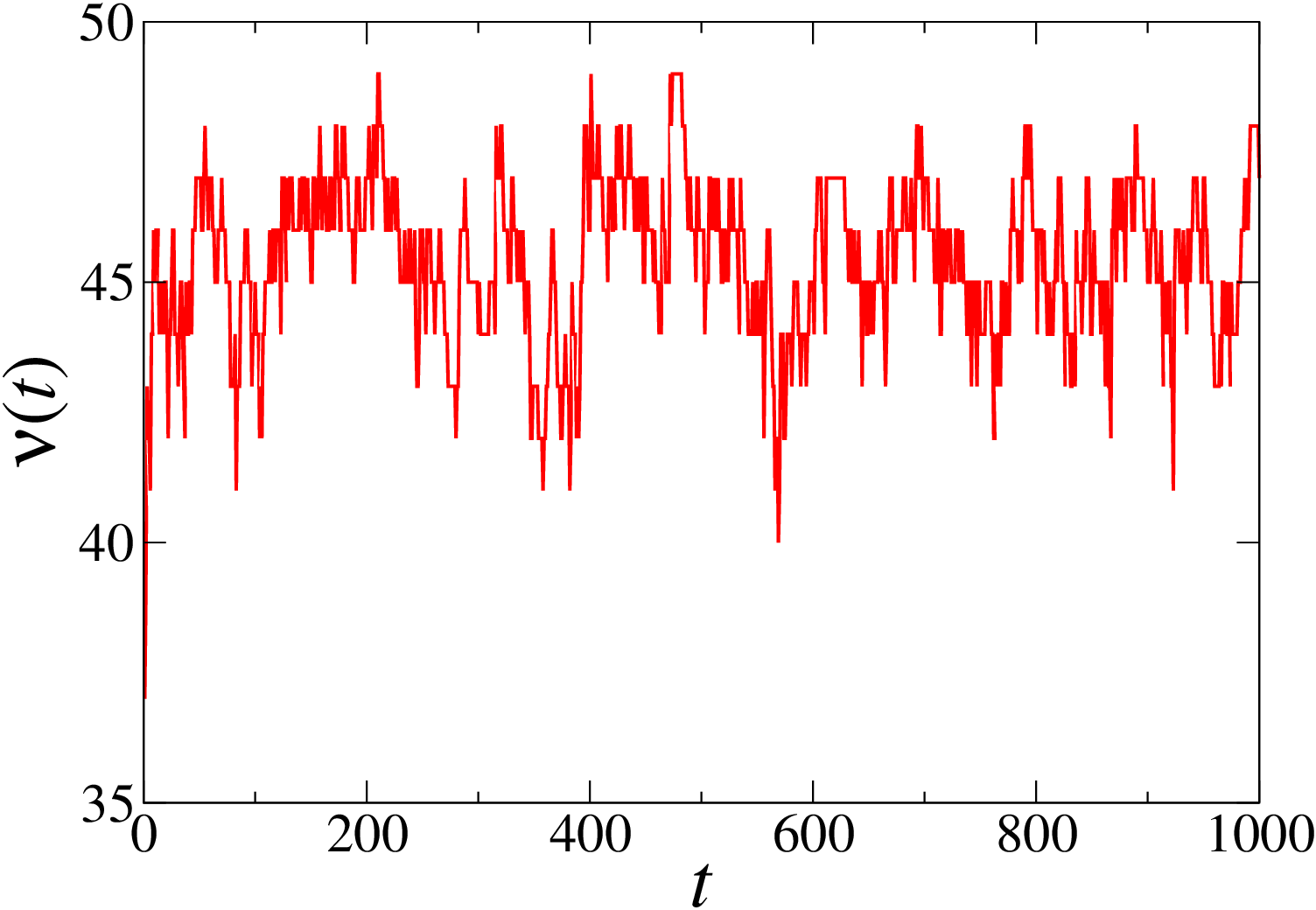}
\caption{\small
Number $\nu(t)$ of spins aligned along their total field
in a typical trajectory where the system is getting lost ($\ell=0.5$, $g=2$, $L=50$).}
\label{wander}
\end{center}
\end{figure}

In spite of its formal zero-temperature form,~(\ref{dylong}) does not represent descent dynamics
any more, because reciprocity is broken in the expressions~(\ref{etalong}) or~(\ref{etazero})
of the total field $\eta_i$.\footnote{The idea that non-reciprocal interactions
might induce non-trivial dynamical effects in spin models,
such as magnetisation oscillations, seemingly appears first in~\cite{sr}.}
In the present setting, this opens up the possibility of getting lost,
in the sense that the dynamics does not reach any fixed point, and instead wanders endlessly.
This is illustrated in Figure~\ref{wander}, showing the time dependence of the number
\beq
\nu(t)=\#\{i\mid\s_i\eta_i>0\}
\eeq
of spins aligned along their total field at time $t$
in a typical trajectory where the system gets lost.
Whenever the dynamics reaches an attractor,~$\nu(t)$ reaches its maximal value $L$,
and stays put there, in a time $T_0$ much shorter than the plotted range.
Notice that $\nu(t)$ is only slightly below~$L$,
so that the system is only weakly lost, in the sense that
it always wanders in the vicinity of an attractor, reminiscent of a
Flying Dutchman who never strays far from the coasts.

These fluctuating trajectories which fail to settle on an attractor are due to
the novel feature of the non-Hamiltonian term in the driving field.
We expect that this will reflect in some way in the retrieval of a heard word, and, in the latter context,
can be thought of as the intermittency associated with the jammed phases of glassy or granular systems~\cite{uspnas}.

\subsection{Phase diagram}
\label{longphase}

The novel features of our composite model proposed for long words
are rooted in the competition between the two components of the total field~$\eta_i$
(see~(\ref{etalong})), i.e., the short-ranged conservative part and the non-local directed one.
These together lead to the possibility of getting lost in word decryption.
The coupling constant~$g$ measures the relative strength of the non-local component.
The following two limit regimes are therefore to be expected:

\begin{itemize}

\item
{\it Disordered regime.}
At small $g$, random local interactions dominate,
so that the picture is qualitatively similar to that of the model studied in Section~\ref{typical}.
There are exponentially many attractors,
and the typical time to reach or to recover any one of them is nearly instantaneous, growing as $\ln L$.

\item
{\it Ferromagnetic regime.}
This novel regime appears at large $g$, where non-local directed interactions dominate.
Typically, there are only a few attractors, each of them exhibiting ferromagnetic order.
There is a ballistic back-propagation of this long-range order from the end of the word,
so that the mean convergence time (conditioned on not getting lost) grows linearly with $L$.

\end{itemize}

The above picture will be corroborated and made quantitative in the following.
We will demonstrate that the model exhibits both regimes whose main characteristics are as described above,
and that these regimes are separated by a rather abrupt crossover in the $\ell$--$g$ plane.
In the present one-dimensional setting,
this crossover does not become a sharp phase transition in the thermodynamic limit.
The rationale behind this expectation is that the coupling constant $g$ might play, in the present
non-equilibrium situation,
the role of the reduced exchange coupling $\beta J$ in the ferromagnetic Ising chain at thermal equilibrium
(see Appendix~\ref{appferro}).
This qualitative correspondence should hold as long as the interaction range~$\ell$ is finite,
so that the directed interactions are non-local but not long-ranged.
It is here useful to make the following interjection: Ising's pioneering work~\cite{ising} showed
that the ferromagnetic chain only exhibits long-range order at zero temperature,
which corresponds either to $g=\infty$ or to $\ell=\infty$ in the present setting.
The absence of phase transitions in one-dimensional models at thermal equilibrium
with finite-ranged interactions is a rather general result
due to Landau~\cite{LL} (see~\cite{CS} for a modern critical account).

Returning to our model, we remark also that
there are many features in our word decryption problem that resemble those found in
typical NP-hard combinatorial optimisation problems,
such as e.g.~constraint satisfaction and graph colouring problems (see~\cite{HW,MM} for overviews).
The abrupt crossover which we will observe in the $\ell$--$g$ plane can therefore be regarded,
at least qualitatively, as a one-dimensional projection of the phase transitions
(usually a static transition preceded by a dynamical one)
observed in NP-hard optimisation problems, to the present problem of decrypting a finite word.
We will have more to say about this in Section~\ref{disc}.

In order to characterise the crossover between the disordered and ferromagnetic regimes,
we have performed numerical simulations of the model.
In each run, the dynamics~(\ref{dylong}) are imposed on
a disordered initial configuration of spins.
They may then either converge to an attractor, i.e., a valid word $W$, or get entirely lost.
From here on, unless otherwise stated, we set $w=1$ and use random spin updates.

The above-mentioned crossover manifests itself most clearly
in the behaviour of the correlation length $\xi$
describing the exponential fall-off of spin correlations
\beq
\mean{\s_i\s_j}\sim\exp(-\abs{i-j}/\xi)
\eeq
in the ensemble of attractors generated by our algorithm.
The model is expected to be deep in the disordered regime for $\xi\ll L$,
and deep in the ferromagnetic regime for $\xi\gg L$.
We have used the magnetisation kurtosis $B$, defined in~(\ref{bdef}),
as a numerical measure of the correlation length $\xi$.
The quantity $B$ is expected to be robust and insensitive to details of the model, since it is
defined as a dimensionless ratio.
More generally, the kurtosis is known to be a suitable tool to analyse the data produced by
numerical simulations.
We have accordingly extracted estimates of $\xi$ from the finite-size scaling formula~(\ref{bfss}).

\begin{figure}[!ht]
\begin{center}
\includegraphics[angle=0,width=0.9\linewidth,clip=true]{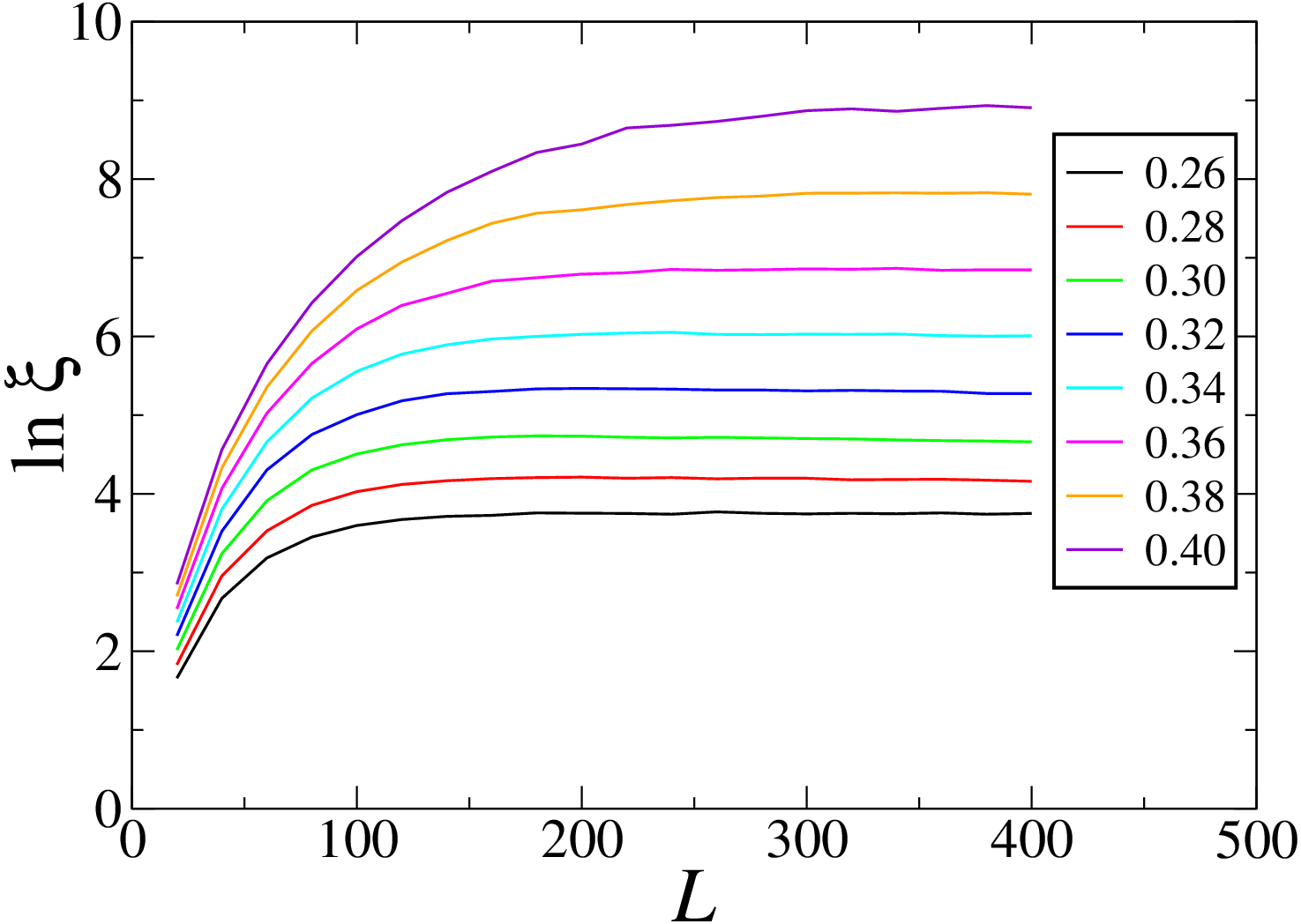}
\caption{\small
Logarithmic plots of the correlation length $\xi$
against $L$ for $\ell=6$ and several~$g$ (see legend).}
\label{xiplot}
\end{center}
\end{figure}

Figure~\ref{xiplot} shows logarithmic plots of the correlation length $\xi$ thus measured on finite samples,
against~$L$ for $\ell=6$ and several $g$.
The data are observed to saturate to well-defined limiting values,
to be identified with the intrinsic correlation length of the model for given $\ell$ and $g$.
The transient regime before saturation is larger for larger~$\xi$, as might be expected.
Note especially that the kurtosis method allows for an accurate determination of correlation lengths $\xi$
that are up to 20 times larger than the sample length $L$.
Last but not least, the correlation length varies over several orders of magnitude in a limited range of $g$.
This clearly manifests the abruptness of the crossover between the disordered and ferromagnetic regimes.

Since the purpose of the present model is to describe long words,
we define a crossover length that is well past the mean word length.
Long words are those whose length $m$ is typically twice the mean word length $\mean{m}$,
i.e., $m\sim20$ (see Table~\ref{abtab}) or $L\sim120$ (see~(\ref{lm})).
Accordingly, we locate the crossover between the disordered and ferromagnetic regimes
by defining the crossover coupling $g_c$ as the value of $g$ such that $\xi=100$.
For $\ell=6$ (see Figure~\ref{xiplot}), this leads to a value of $g_c\approx 0.30$.

\begin{figure}[!ht]
\begin{center}
\includegraphics[angle=0,width=0.9\linewidth,clip=true]{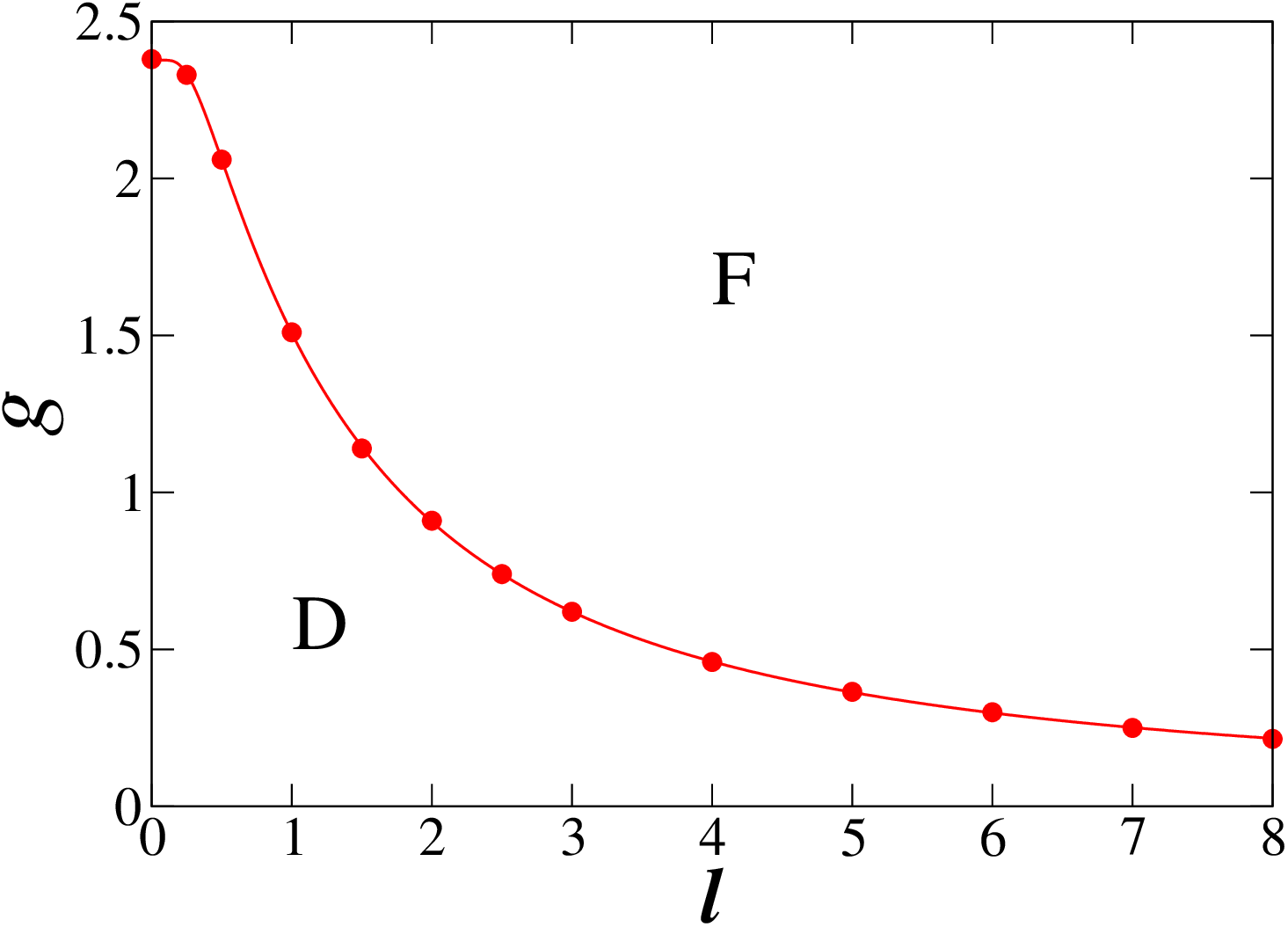}
\caption{\small
Phase diagram in the $\ell$--$g$ plane.
Symbols: numerical data for the crossover coupling $g_c$, defined by the condition $\xi=100$.
Full curve: fit of $g_c$ by a polynomial in $x$ (see text).
D: disordered regime.
F: ferromagnetic regime.}
\label{phase}
\end{center}
\end{figure}

Figure~\ref{phase} shows the ensuing `phase diagram' in the $\ell$--$g$ plane.
The symbols show our numerical estimates for the crossover coupling $g_c$
for several values of the interaction range $\ell$.
The limiting value $g_c\approx2.38$ as $\ell\to0$ has been measured
by means of a direct simulation of the nearest-neighbour model (see~(\ref{etazero})).
This number is in excellent agreement with the approximate prediction derived in
Appendix~\ref{appell} (see~(\ref{gcz})),
suggesting that the latter prediction might become exact in the $\ell\to0$ limit.
The full curve shows a fit of $g_c$ by a fourth-degree polynomial in $x$
(see~(\ref{xdef})) with no constant term.
This representation describes the data accurately,
including its structure at small $\ell$,
and predicts the large-$\ell$ decay
\beq
g_c\approx\frac{C}{\ell},\qquad C\approx1.43.
\eeq
The plot confirms what one might expect: ferromagnetic order sets in even for modest interaction
strengths $g$ for large interaction ranges $\ell$,
while for small interaction ranges, including the case of nearest-neighbour coupling ($\ell\to0$),
interaction strengths need to be larger for ordering to occur.
We will have more to say in Section~\ref{disc} about what this means in the context of real words.

The crossover between the disordered and ferromagnetic regimes manifests itself
in a panoply of static and dynamical quantities.
On the static side, the key quantity is the number $\N$ of attractors.
We recall that the attractors are the fixed points of the dynamics~(\ref{dylong}),
obeying~(\ref{fp}), where the total field $\eta_i$ is given by~(\ref{etalong}).
In the latter equation, the variable-range component of the interactions is totally directed,
so that the recursive computer algorithm used in Section~\ref{typical}
to enumerate all attractors can still be used in the present setting.

\begin{figure}[!ht]
\begin{center}
\includegraphics[angle=0,width=0.9\linewidth,clip=true]{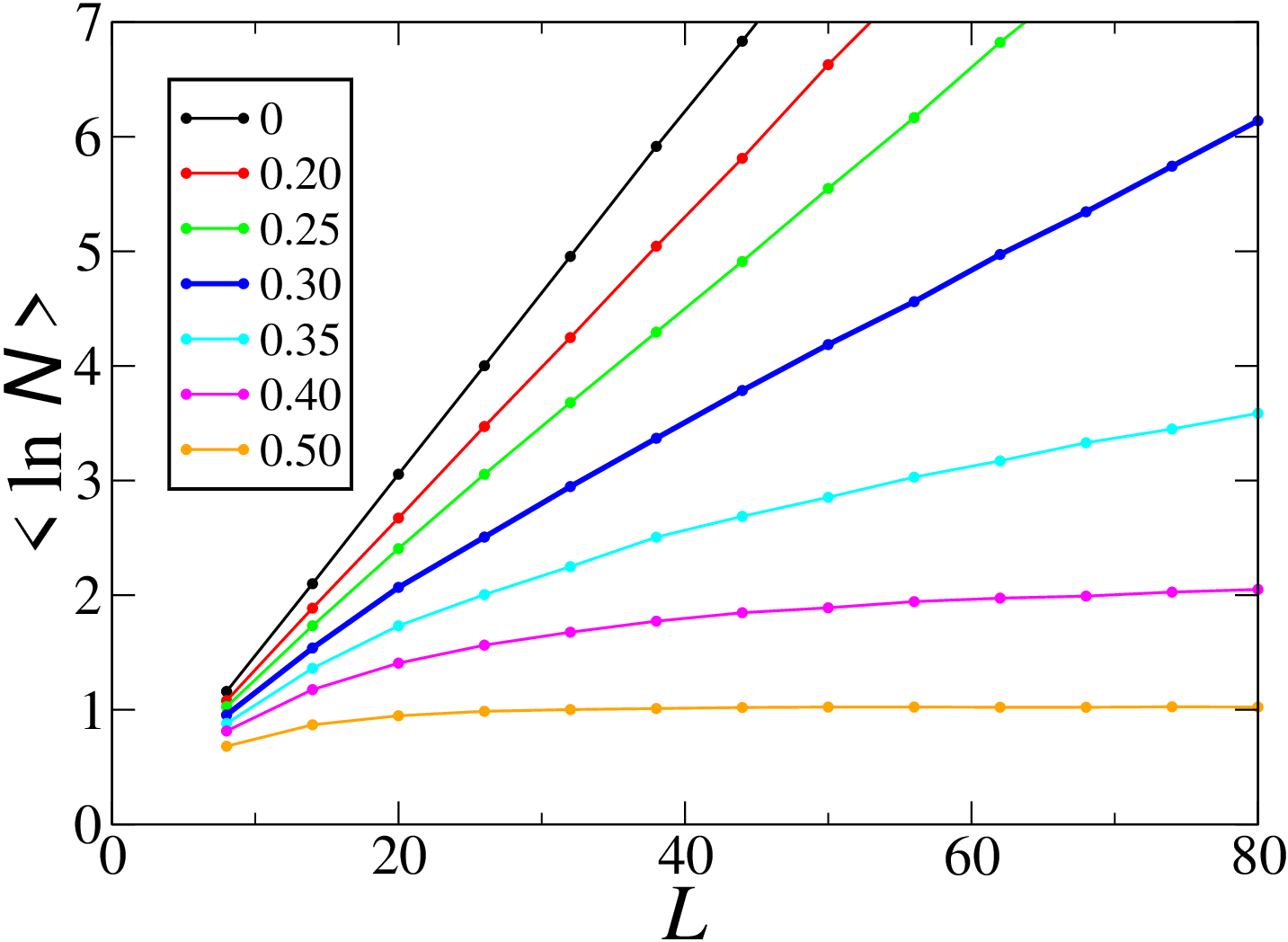}
\caption{\small
Plot of $\mean{\ln\N}$ against $L$ for $\ell=6$ and several~$g$ (see legend).
The thick blue curve shows data for $g=0.30\approx g_c$.}
\label{slog}
\end{center}
\end{figure}

Figure~\ref{slog} shows a plot of $\mean{\ln\N}$ against $L$
for $\ell=6$ and several values of the coupling constant~$g$.
The data have been obtained by means of numerical simulations
based on the recursive algorithm mentioned above.
The plotted quantity $\mean{\ln\N}$ grows linearly in $L$ at small coupling, where $\xi\ll L$,
in agreement with the exponential growth law~(\ref{nexp}),
whereas it seems to asymptote to finite values at large coupling, where $\xi\gg L$.
These results confirm our expectations: we would expect a great many small words, while
longer words would be progressively fewer.
The dependence of $\mean{\ln\N}$ on $g$ is at its most sensitive
in the crossover regime near $g_c\approx0.30$ (thick blue line).

\begin{figure}[!ht]
\begin{center}
\includegraphics[angle=0,width=0.9\linewidth,clip=true]{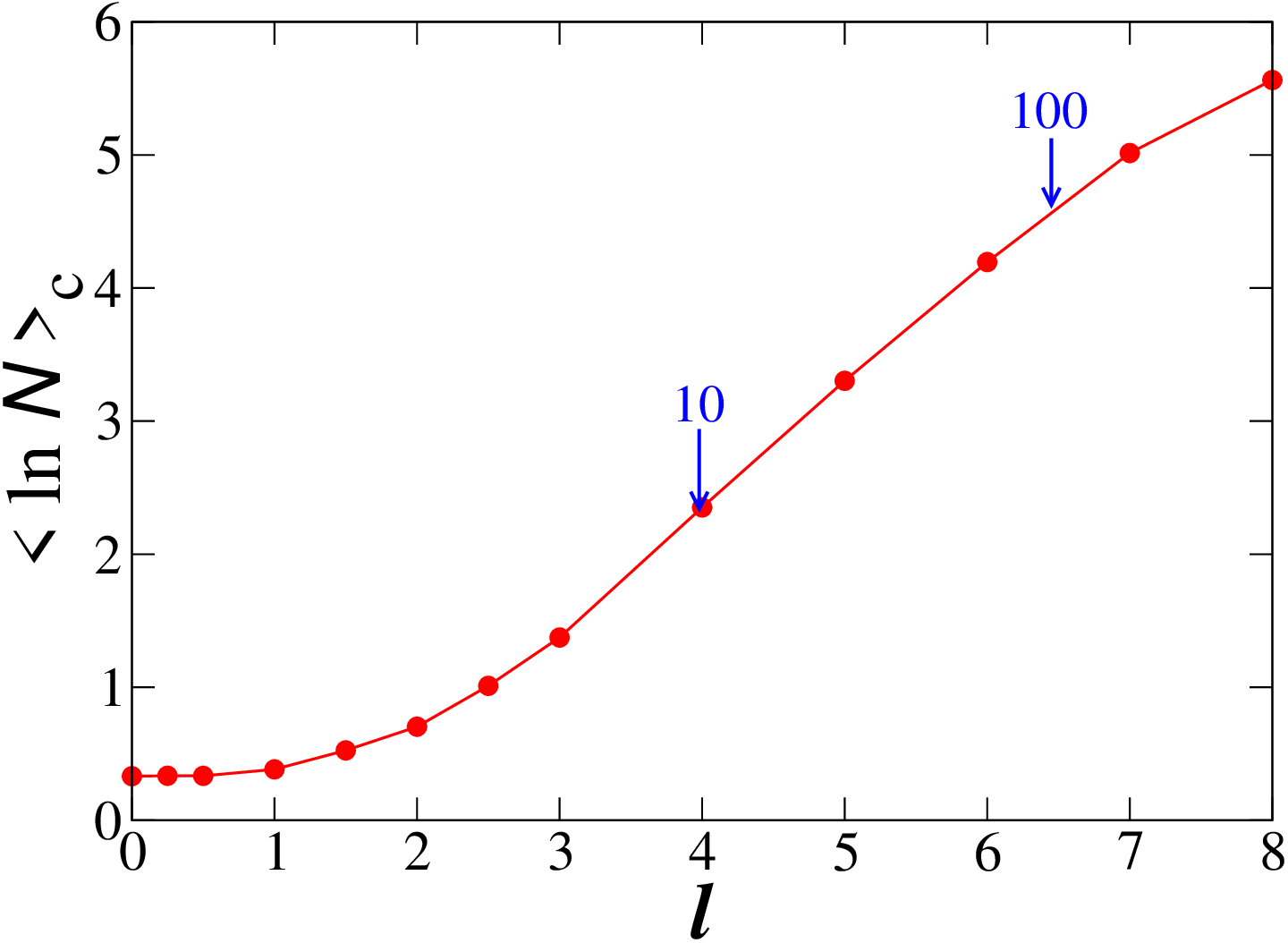}
\caption{\small
Plot of $\mean{\ln\N}_c$, measured right at the crossover coupling $g=g_c$
for $L=50$ and the same sequence of values of $\ell$ as in Figure~\ref{phase}.
Arrows: values of $\ell$ such that the typical number of attractors,
$(\N_\typ)_c=\exp\mean{\ln\N}_c$, equals 10 and 100.}
\label{scrit}
\end{center}
\end{figure}

Figure~\ref{scrit} shows a plot of $\mean{\ln\N}_c$,
measured right at the crossover coupling $g=g_c$,
for $L=50$ and the same sequence of values of $\ell$ as in Figure~\ref{phase}.
The typical number of attractors at $g_c$,
i.e., $(\N_\typ)_c=\exp\mean{\ln\N}_c$ (see~(\ref{ntypdef})), varies over a very broad range.
It remains microscopic for interaction ranges $\ell$ up to 3,
before it rises steeply for larger values of $\ell$.
The arrows in the figure show the values of $\ell$ such that the typical number of attractors equals 10 and 100.

\begin{figure}[!ht]
\begin{center}
\includegraphics[angle=0,width=0.9\linewidth,clip=true]{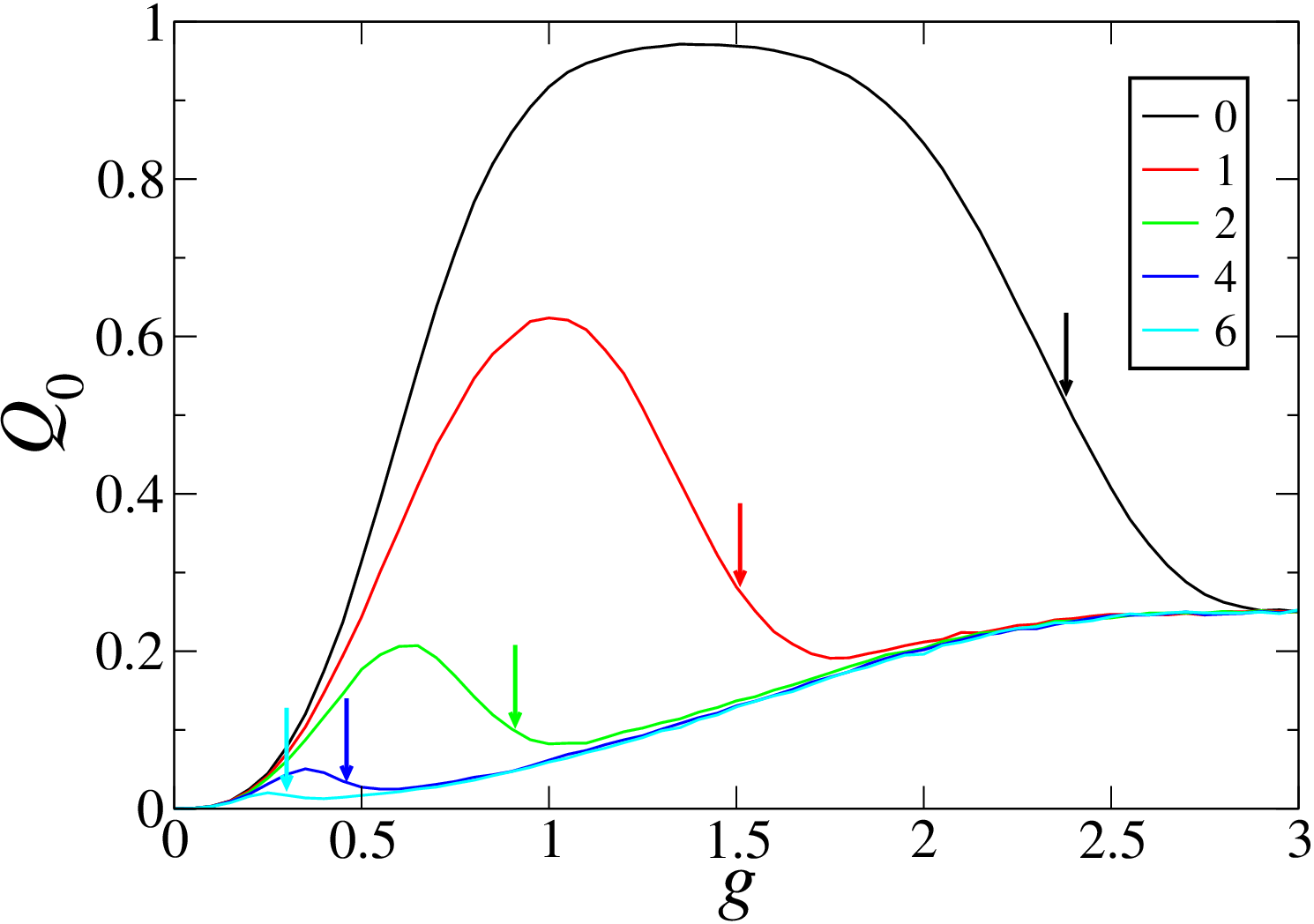}
\caption{\small
Probability $Q_0$ of getting lost
during the search of a valid word from a disordered initial configuration,
plotted against $g$
for $L=100$ and several values of $\ell$ (see legend).
Arrows with corresponding colours: crossover couplings~$g_c$.}
\label{q0}
\end{center}
\end{figure}

As far as the dynamical manifestations of the crossover between the disordered and ferromagnetic
phases are concerned, the first and foremost effect is the possibility of getting lost,
so that the dynamics~(\ref{dylong}) launched from a random initial configuration wanders endlessly,
as illustrated in Figure~\ref{wander}.
Figure~\ref{q0} shows a plot of the probability~$Q_0$ of getting lost
during the search for a valid word from a disordered initial configuration,
against the coupling constant $g$,
for $L=100$ and several values of the interaction range~$\ell$.
The arrows show the corresponding crossover coupling~$g_c$.
At fixed~$\ell$,~$Q_0$ exhibits a maximum at some $g$ below $g_c$,
and a minimum at some $g$ above~$g_c$.
The maximum is more pronounced for small $\ell$.
It reaches the very high value of 0.971 for the nearest-neighbour model,
corresponding to the $\ell\to0$ limit.
All datasets reach the limiting value $Q_0=P_0=1/4$ at $g_0=3$ (see Appendix~\ref{appg}),
beyond which the probability $Q_0$ remains constant.

The key quantities of the full model thus exhibit a very strong dependence on the interaction range $\ell$.
This is the case for the crossover coupling $g_c$ itself (see Figure~\ref{phase}),
for the typical number of attractors at $g_c$, namely $(\N_\typ)_c=\exp\mean{\ln\N}_c$ (see Figure~\ref{scrit}),
and for the probability $Q_0$ of getting lost during the first stage of the dynamics (see Figure~\ref{q0}).
The regime of small interaction ranges, say $\ell<3$,
with its huge probability of getting lost $(Q_0)_c$ and its very small number of attractors $(\N_\typ)_c$ at $g_c$,
seems somewhat unsuitable for modelling word decryption in any realistic regime.
From now on, therefore, we set the interaction range to be
\beq
\ell=6.
\label{ell6}
\eeq
It is significant that this choice of $\ell$ coincides with the length of a string of spins
encoding a single sound or phoneme (see~(\ref{lm})):
in terms of sounds, the interaction range has been set to unity.
For this specific parameter value, we have $g_c\approx 0.30$, $(\N_\typ)_c\approx66.4$ and $(Q_0)_c\approx0.0169$.

\begin{figure}[!ht]
\begin{center}
\includegraphics[angle=0,width=0.9\linewidth,clip=true]{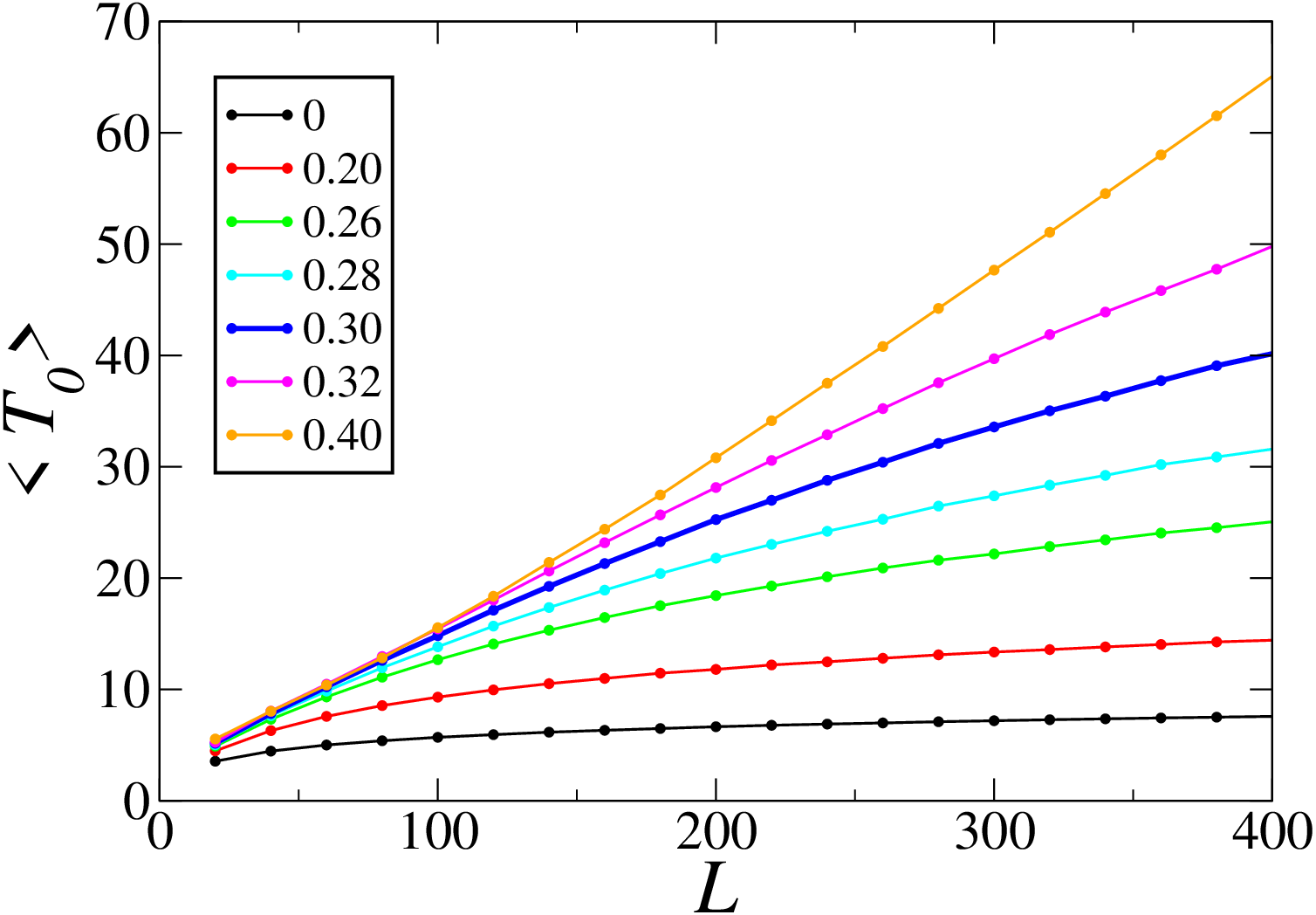}
\caption{\small
Mean time $\mean{T_0}$ taken
to select a word $W$ from a disordered initial configuration,
plotted against $\ln L$ for $\ell=6$ and several $g$ (see legend).
The thicker blue curve shows data for $g=0.30\approx g_c$.}
\label{tdyn}
\end{center}
\end{figure}

The next dynamical quantity of interest is the time $T_0$ taken by the dynamics to select a valid word $W$
from a disordered initial configuration, i.e., to converge to an attractor.
Figure~\ref{tdyn} shows a plot of $\mean{T_0}$ against $L$
for $\ell=6$ and several values of the coupling constant~$g$.
The data have been obtained by means of numerical simulations of the dynamics~(\ref{dylong}).
Histories where the system wanders forever, so that $T_0=\infty$,
are discarded from the ensemble of histories used to define $\mean{T_0}$.
At small coupling, this quantity grows very slowly with $L$,
in agreement with the logarithmic law~(\ref{tlog}) observed at $g=0$.
At strong coupling, the linear growth of $\mean{T_0}$
agrees with the picture of ballistic ordering.
Finally, the dependence of $\mean{T_0}$ on~$g$ is again the strongest near $g_c\approx0.30$ (thick blue line).
It is instructive to compare this plot with Figure~\ref{slog}, for a full interpretation of our results.
There, the thick blue line ($g_c\approx0.30$)
separates regions where attractors are abundant ($g<g_c$) from those where they are rare ($g>g_c$).
Clearly, one should expect
that the dynamics will be swift (i.e., the time $T_0$ to arrive at an attractor will be small)
where attractors are abundant ($g<g_c$).
Equally, as valid words become rare ($g>g_c$), i.e., attractors are sparser,
the dynamics will be increasingly slow.
This is precisely what is observed in Figure~\ref{tdyn}, on either side of the thick blue line.

We end this section by describing explicitly the connection between the parameters
of the present model and realistic word length distributions.
Using the proportionality~(\ref{mrn}) between the number $n$ of letters in a given word
and the number $m$ of its sounds,
the Gamma distribution~(\ref{gamma}) can be recast as
\beq
\widehat{p}_m=\widehat{C}\,m^\alpha\,\e^{-\widehat{\beta}m},
\label{mgamma}
\eeq
with $\widehat{\beta}=\beta/r$.
If the total lexicon size is~$\Lambda$,
the number of words consisting of $m$ sounds is therefore estimated as $\Lambda\widehat{p}_m$.
A quantitative correspondence with our model can be established as follows.
For a given word length $m$, and a given interaction range $\ell$,
we define an effective coupling~$g_\eff$ as the value of~$g$ such that
\beq
\mean{\ln\N}=\ln(\Lambda\widehat{p}_m).
\eeq

\begin{figure}[!ht]
\begin{center}
\includegraphics[angle=0,width=0.9\linewidth,clip=true]{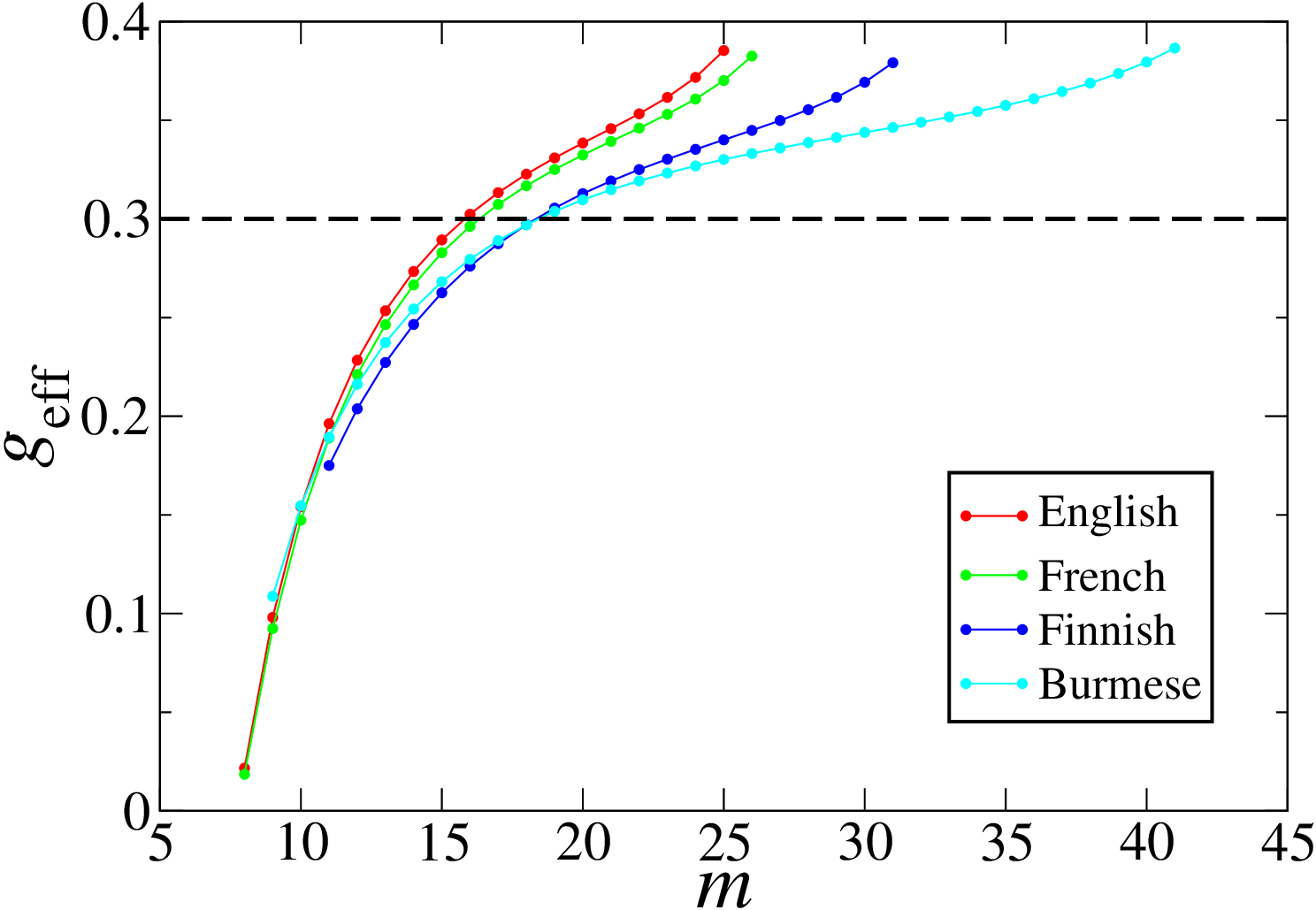}
\caption{\small
Effective coupling $g_\eff$ plotted against $m$ for $\ell=6$, $\Lambda=10^5$,
and four languages (see legend), whose parameters are given in Table~\ref{abtab}.
Dashed line: crossover coupling $g_c\approx0.30$.}
\label{wld}
\end{center}
\end{figure}

Figure~\ref{wld} shows the effective coupling~$g_\eff$ thus defined plotted against $m$
with $w=1$, $\ell=6$, $\Lambda=10^5$,
for the four languages whose parameters are given in Table~\ref{abtab}.
Only word lengths for which more than ten words are predicted by~(\ref{mgamma}) are shown in the figure.
As might be expected, $g_\eff$ virtually vanishes for words
whose length equals the mean word length $\mean{m}$;
it then increases steadily as $m$ is increased beyond $\mean{m}$,
and reaches the crossover coupling $g_c$ (dashed line)
for word lengths which are about twice the average.
It then increases more rapidly deeper in the tail of the word length distribution.
The above equivalence provides a useful framework
for comparing the predictions of the present model with existing linguistic data.

\begin{table}[!ht]
\begin{center}
\begin{tabular}{|l|c|c|c|c|c|}
\hline
Language & $\alpha$ & $\beta$ & $\widehat{\beta}$ & $\mean{n}$ & $\mean{m}$ \\
\hline
English & 4.4 & 0.60 & 0.71 & 8.3 & 7.1 \\
French & 4.9 & 0.60 & 0.71 & 10.1 & 8.6 \\
Finnish & 6.8 & 0.58 & 0.68 & 13.4 & 11.4 \\
Burmese & 2.6 & 0.28 & 0.33 & 12.8 & 10.9 \\
\hline
\end{tabular}
\vskip 6pt
\caption{
Numerical values of the parameters $\alpha$, $\beta$ and $\widehat{\beta}=\beta/r$
entering the forms~(\ref{gamma}) and~(\ref{mgamma}) of the length distributions of written and
spoken words,
and resulting average lengths $\mean{n}$ and $\mean{m}=r\mean{n}$,
for the four languages shown in Figure~\ref{wld}.
Values of $\alpha$, $\beta$ and $\mean{n}$ are taken from~\cite{us}.}
\label{abtab}
\end{center}
\end{table}

\subsection{Decrypting misheard long words}
\label{longdecrypt}

We now investigate the task of decrypting a mispronounced or misheard long word.
The decryption task is still modelled as in~(\ref{scheme}),
with the following essential difference.
For typical words, the descent dynamics~(\ref{dy}) involves the local field $\eta_i$ defined in~(\ref{eta}),
whereas, in the present case of long words,
the local field~(\ref{etalong}) comprises a symmetry-breaking term.
Thus we are no longer dealing with a {\it pure} descent dynamics.
In particular, this situation results in the possibility of getting lost during the decryption stage.

\begin{figure}[!ht]
\begin{center}
\includegraphics[angle=0,width=0.9\linewidth,clip=true]{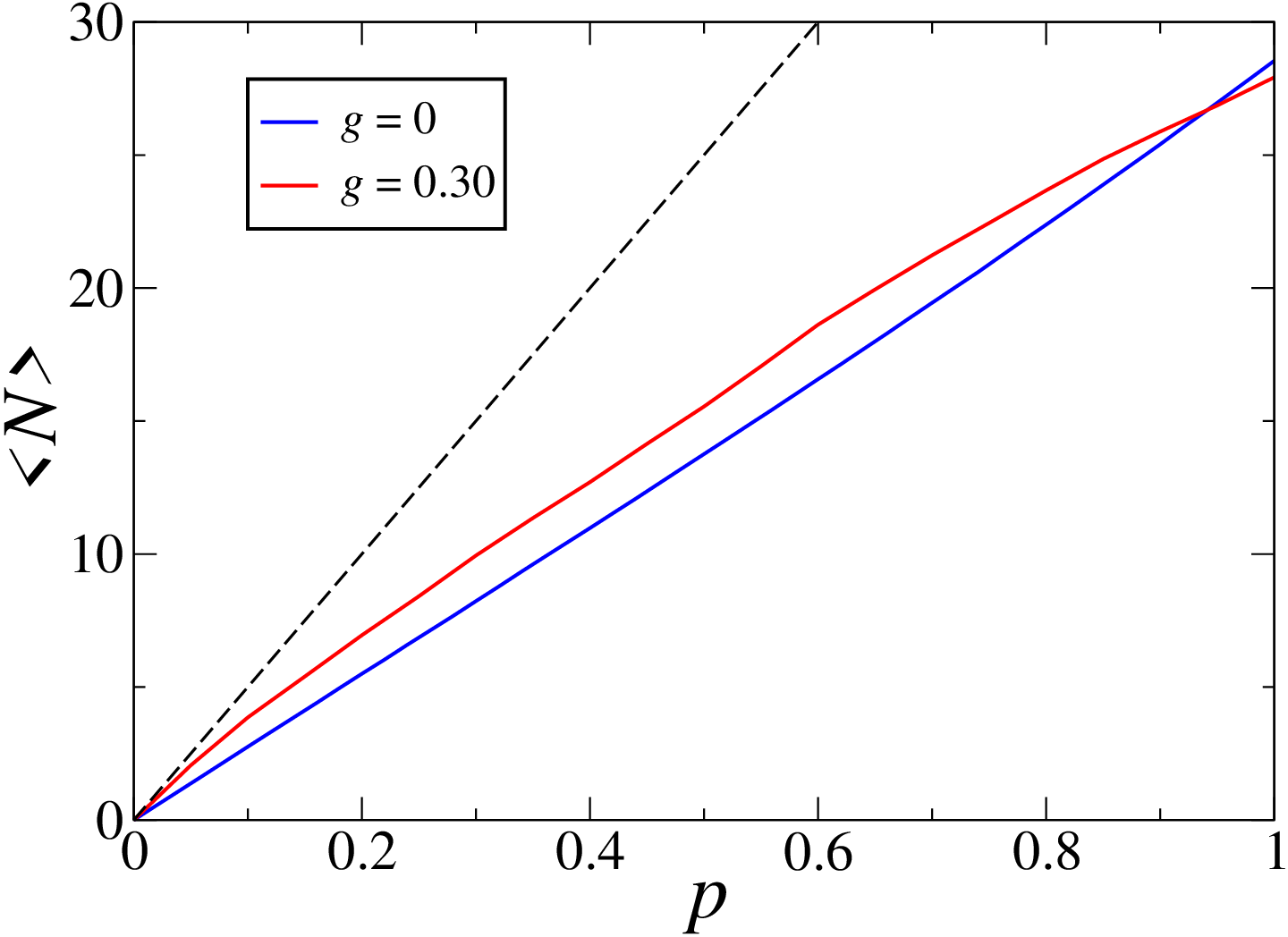}
\caption{\small
Mean number $\mean{N}$ of residual mishearings against density $p$ of initial mishearings,
for a random spin update
with $\ell=6$, $L=50$ and two values of the coupling constant, namely $g=0$ and
$g=0.30\approx g_c$
(see legend).
Black dashed line: $\mean{N_0}$ (see~(\ref{nzero})).}
\label{longave}
\end{center}
\end{figure}

\begin{figure}[!ht]
\begin{center}
\includegraphics[angle=0,width=0.9\linewidth,clip=true]{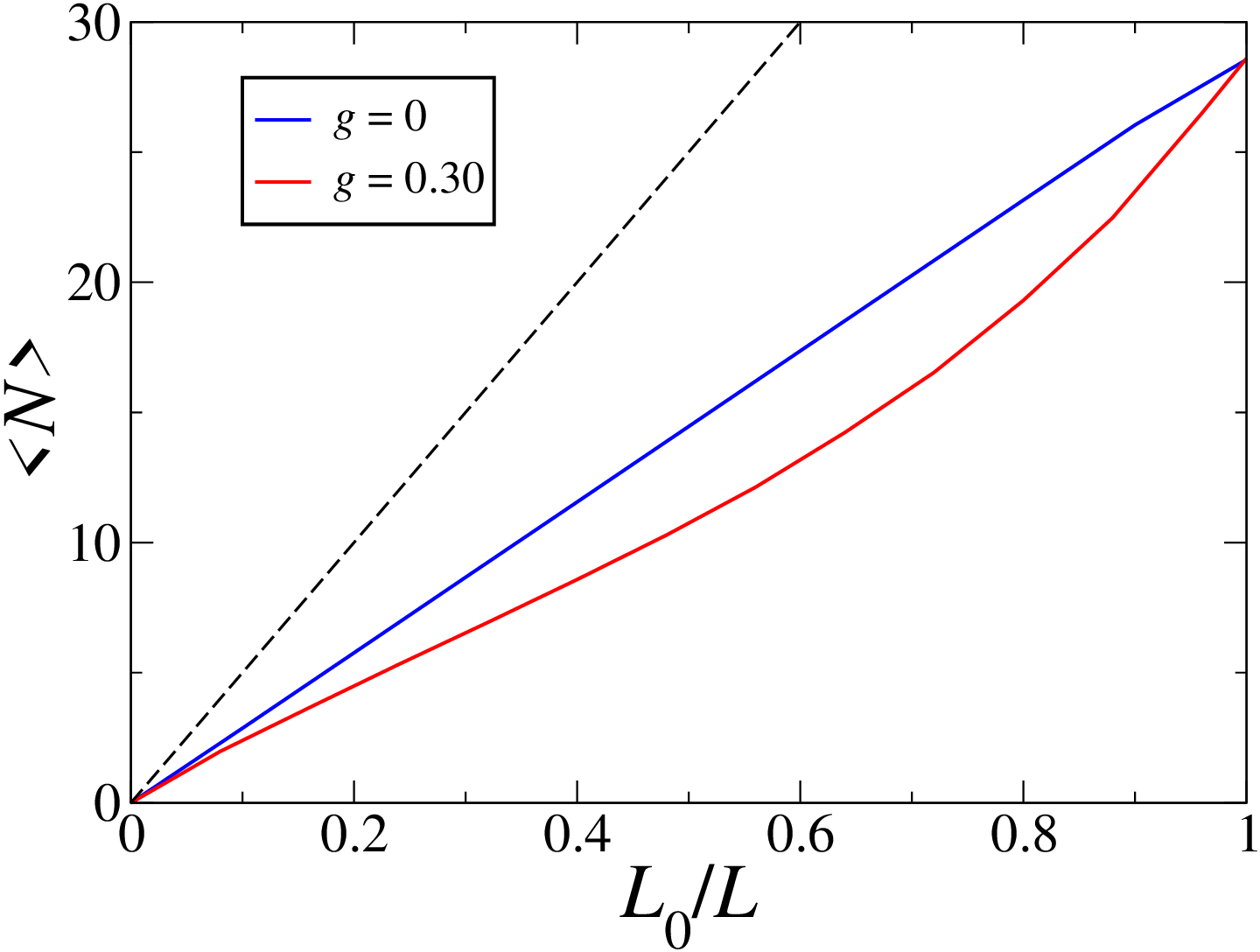}
\caption{\small
Mean number $\mean{N}$ of residual mishearings against initial fraction $L_0/L$ of clustered
mishearings,
for the same parameter values as in Figure~\ref{longave}.
Black dashed line: $\mean{N_0}=L_0$.}
\label{longcluster}
\end{center}
\end{figure}

Figure~\ref{longave} shows a plot of the mean number~$\mean{N}$ of residual mishearings
against the density $p$ of initial mishearings for a random spin update
with $\ell=6$, $L=50$, and two values of the coupling constant, namely $g=0$ and $g=0.30\approx g_c$.
The first dataset therefore coincides with the first dataset of Figure~\ref{ave}.
We have also run simulations of the decryption model in the situation of clustered mishearings;
here, instead of being distributed independently and uniformly along the word $W$,
all mishearings are {\it consecutive} and occupy a central portion of length $L_0$.
Assuming for definiteness that $L$ and $L_0$ are both even,
the stochastic rule~(\ref{sigmap}) is thus replaced by the following deterministic one:
\beq
\s_i^\Sigma=\left\{
\begin{matrix}
\hfill\s_i^W\;&\hbox{for}\;1\le i\le\half(L-L_0),\hfill\cr
\hfill-\s_i^W\;&\hbox{for}\;\half(L-L_0)<i\le\half(L+L_0),\hfill\cr
\hfill\s_i^W\;&\hbox{for}\;\half(L+L_0)<i\le L.\hfill\cr
\end{matrix}
\right.
\eeq
The corresponding data for the mean number $\mean{N}$ of residual mishearings after decryption
are plotted in Figure~\ref{longcluster} against the initial fraction of clustered mishearings,
i.e., the ratio $L_0/L$, for the same parameter values.
The small fraction of instances where the system gets lost during the decryption stage of the dynamics
are discarded from the datasets shown in Figures~\ref{longave} and~\ref{longcluster}.

The comparison between the two above figures
demonstrates that the addition of symmetry-breaking interactions
slightly reduces the efficiency of the decryption algorithm
in the case of a uniform density of mishearings (Figure~\ref{longave}),
while it improves the efficiency
in the case of clustered mishearings (Figure~\ref{longcluster}).
Although modest, this improvement is more significant for higher values of the ratio $L_0/L$.
The two opposite effects described here take place for generic values of the coupling constant $g$;
they are not limited to the vicinity of the crossover coupling $g_c$, and originate in the fact
that an increase of $g$ implies a stronger `back-reaction' from the end of the word.
In the case of clustered mishearings (Figure~\ref{longcluster}), this back-reaction has the effect of `pinning'
the end of a partially decoded and therefore more ordered cluster (especially if it is long).
This immediately leads to an increased efficiency.
No such effect occurs for randomly placed mishearings
(Figure~\ref{longave}), where the effect of an increased $g$ might lead to the pinning of the end
of the word, despite the presence within it of randomly placed mishearings.

\begin{figure}[!ht]
\begin{center}
\includegraphics[angle=0,width=0.9\linewidth,clip=true]{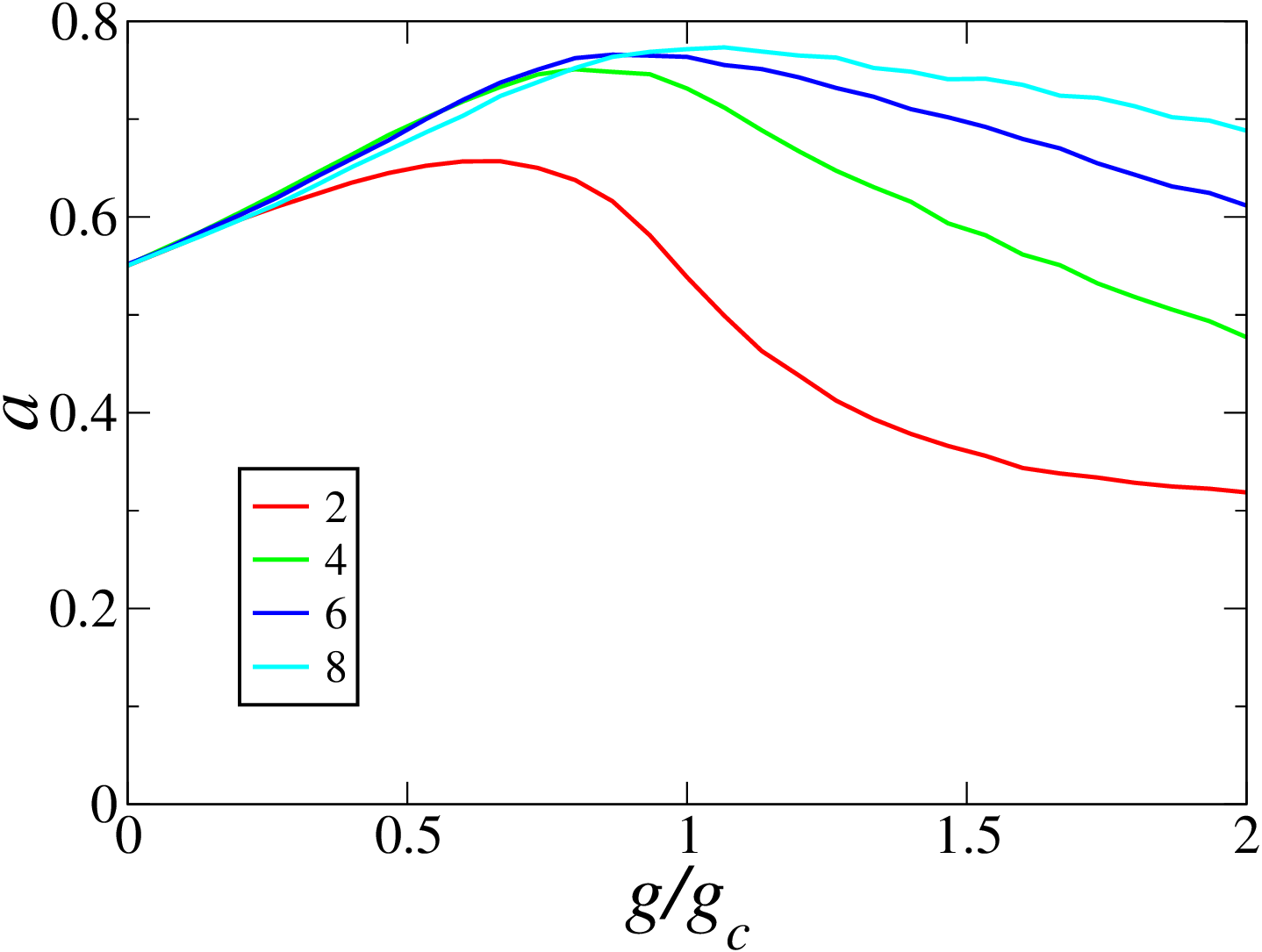}
\caption{\small
Amplitude $a$ characterising the efficiency of the decrypting algorithm
in the regime of dilute mishearings,
plotted against the ratio $g/g_c$ for several values of the interaction range $\ell$ (see legend).}
\label{along}
\end{center}
\end{figure}

To end this analysis,
we consider again the scaling regime where the system length $L$ is large,
whereas the density $p$ of mishearings is small.
In this regime, the mean number of residual mishearings
still scales as $\mean{N}\approx a p L$ (see~(\ref{npsca})).
The amplitude~$a$, providing a quantitative measure of the efficiency of the decrypting algorithm,
is plotted in Figure~\ref{along} against the ratio $g/g_c$ for several values of the interaction range $\ell$.
All plots start from the value $a\approx0.552$ (see Figure~\ref{abplot}, for random spin updates).
They seem to have the same initial slope, irrespective of $\ell$; this would seem to be reasonable
because until the coupling $g/g_c$ is significant,
one would not expect to see the effect of the interaction range $\ell$.
They reach a smooth maximum, corresponding to the poorest efficiency, in the vicinity of $g_c$.
Beyond this crossover coupling, the number of words decreases rapidly,
and the constraints (`back-reactions') from the end of the word increase,
leading to a sharp decrease in the amplitude~$a$.
These concomitant effects are more and more pronounced as the interaction range $\ell$ is decreased.

\section{Discussion}
\label{disc}

The main purpose of this paper is to present a microscopic model of word recognition.

In earlier work~\cite{us}, we had presented some phenomenological ideas on how to construct words,
and what the mechanisms of their retrieval might be, with or without the presence of mishearings.
That work was essentially based on estimating the numbers of words, word variants,
and word prefixes of prescribed lengths.
These `phase-space' arguments were successful in describing some of the static aspects of the problem
(e.g.~the mean number of sounds needed to anticipate a spoken word),
but were inadequate for a representation of the dynamics of word recognition.
A major reason for this was that we had in that work recognised, but not implemented,
the importance of correlations between sounds along any word.
In this paper, we are able to construct a more microscopic model of word recognition
based on an algorithmic and dynamical approach,
which incorporates the needed correlations in a simple and natural fashion.

First, and in consonance with realistic word length distributions,
such as e.g.~the Gamma distribution~(\ref{gamma}),
our algorithm differentiates between the recognition of short and long words:
short words, whose length sits either before or at the peak of the word length distribution, are abundant,
while longer words are increasingly rare as a function of their length.
In practice this leads to the relatively `easy' recognition of a short word, even in the presence of mishearings.
For long words, the decryption of a misheard word soon becomes more complex.

Accordingly, we model the speech recognition process as some descent dynamics on a spin chain,
where the total aligning field comprises two parts.
The first part derives from a short-ranged, disordered spin Hamiltonian,
leading to a multiplicity of attractors which is adequate to describe short words.
The second part of the aligning field does not derive from a Hamiltonian.
It is totally directed, thus breaking reciprocity and detailed balance.
This symmetry-breaking component is characterised by a coupling constant $g$ and an interaction range $\ell$,
which constitute the two main parameters of the present model.
The full setting, with its two components,
is adequate to model the recognition (or not) of misheard words, irrespective of their length.

The actual decryption process is chosen to be the zero-temperature single-spin-flip dynamics
induced by the above aligning field.
The fixed points of the dynamics correspond to the words of our lexicon.
A misheard word is therefore progressively `unravelled' via these dynamics until a plausible word is reached.
For short words, this goes through smoothly,
with expected features such as ordered updates being faster than random updates in word retrieval.
This is because the ordered sequential algorithm essentially shaves off mishearings as it proceeds,
almost never going back to them as it progresses to word recovery.

The novelty of the present model really surfaces when longer words
(i.e., those sitting significantly beyond the peak of the word length distribution)
are decrypted.
Here, we encounter for the first time the phenomenon that some misheard words are never recognised.
Instead, the algorithm hovers endlessly as it tries to retrieve a misheard word,
constantly sampling possible words in the neighbourhood of the original (mis)heard word.
This possibility of getting lost is a central result of this paper,
and a firm indicator that a level of dynamical complexity has been achieved.
It is of course all too real, in practice, to be lost when trying to decipher misheard speech.

The detailed analysis of the attractors of the descent dynamics
reveals an abrupt crossover in the $\ell$--$g$ plane
between a `disordered' and a `ferromagnetic' regime.
The crossover takes place at some `critical' coupling $g_c$, depending on the interaction range $\ell$.
Its main static manifestation is the sudden appearance of long-range ferromagnetic order
among the spins describing a typical attractor, i.e., a typical word with appropriate length.
The disordered regime ($g<g_c$) roughly corresponds to short words,
whose length is close to typical.
These words are abundant,
and their algorithmic decryption is fast and relatively easy, although imperfect.
The ferromagnetic regime ($g>g_c$) corresponds to long words,
whose lengths are significantly larger than typical.
These words are less abundant,
and their algorithmic decryption becomes progressively slower and less easy.
The more direct dynamical implication of this crossover is that
the probability of getting lost is maximal for a subcritical coupling $g\approx2g_c/3$ (see Figure~\ref{q0}).
We have also established the explicit correspondence between the parameters of our present model
and actual word length distributions for four languages analysed in our previous work~\cite{us}.
In agreement with the above,
the crossover is observed to take place for words whose lengths are roughly twice as long as average,
irrespective of the language under study.

At this point it is useful to explore in some more detail the role of dimensionality in the problem.
Recall that the real task of decrypting words that may well be long and complex,
is one whose dimensionality is unclear;
for example non-local moves in search space clearly imply dimensionalities higher than one.
Also, word decryption is likely to be an NP-hard problem.
Many such problems, such as e.g.~constraint satisfaction and graph colouring problems,
are defined on complete graphs or cliques, pertaining to an infinite-dimensional or mean-field
geometry; they are known to exhibit phase transitions, and their algorithmic complexity
may have a subtle dependence on spatial dimensionality.\footnote{To take one example,
finding the ground-states of an Ising spin glass
is an easy problem in two dimensions, namely on planar graphs~\cite{gre},
and a NP-hard one in higher dimensions.
The problem is already hard in the quasi-two-dimensional geometry
of two planes coupled by transverse bonds
(see~\cite{KW} and references therein).}
In this paper, we have made a transcription of the real problem onto a spin chain,
imitating the flow of speech over time.
Since only local moves are allowed in our model, it is one-dimensional in essence.
The advantage of this simplified description is that it enables an easier analysis of the original,
high-dimensional word decryption problem,
and the presentation of results which are relevant, at least qualitatively, to the original problem.
What must, however, be borne in mind is that this reduction to one dimension substitutes a crossover
for any kind of phase transition in the original problem.
The abrupt crossover we see here, between the disordered and ferromagnetic regimes,
thus appears as a projection of possible phase transitions in
the real, possibly infinite-dimensional, problem of word recognition.

Another consequence of the fact that our model is a simplification of the original problem
is that we get a fast one-shot, imperfect answer to word recognition,
i.e., we either land on the intended word, on another valid word, or get lost forever.
Again, in real life, the process of word recognition is more complex.
We tend to turn multiple variants of the misheard words around in our head,
according to the phenomenon called reverberation.
We may choose, for example, to change a single sound, and examine the space of possible words, or to
change small groups of neighbouring sounds simultaneously to resemble a known word (for instance,
the decoding of the heard `hambag' to `handbag' in English~\cite{lahiri1}).
This complex search gives rise to longer periods of reflection, eventually leading to success.
In physical terms, by changing to a richer, more realistic level of description,
we would get a distribution of retrieval times ranging from fast to slow and then slowest.
The absence of such collective effects in the present model causes the behaviour of, say, decryption times,
to not distinguish between a string of isolated mishearings and a cluster of consecutive ones.
It gives a fast answer of success or failure, in a time growing at most ballistically,
i.e., proportionally to the word length, without a broad distribution of long retrieval times.
Therefore, any attempt to find a
dynamical transition in the current model is doomed to failure.
We are currently working on a model
of speech perception which addresses these questions, among others~\cite{us1}.

We also briefly address the issue of what `success' really means in the context of word retrieval in the present model.
We have defined success as the retrieval of the word intended by the speaker.
Failure therefore corresponds to two situations: we
either converge to a different valid word, or we get lost.
In the regime of short words, the neighbourhood of a given word is dense with valid words
so the attractor may well be a word that is significantly different from the original word.
In the regime of long words, on the other hand, attractors soon become rare.
With the understanding that the search process involves finding the end point of a word
with a well-established prefix, one can imagine the retrieved word is rather closely related to the original
(e.g.~`circumspectly' or `circumspection' as possible endpoints of the string `circumspect-').
We envisage, thus, closely spaced attractors residing in a corner of search space.
This is exactly what occurs in a limiting case of our model, where the coupling constant is very large
(see Appendix~\ref{appg}).

\subsubsection*{Acknowledgments}

\small
It is a pleasure for us to thank Guillaume Jacques and Pierfrancesco Urbani for stimulating discussions.
AM warmly thanks the Leverhulme Trust for the Visiting Professorship that funded part of this research,
as well as the Faculty of Linguistics, Philosophy and Phonetics, Oxford
and the Institut de Physique Th\'eorique, Universit\'e Paris-Saclay, for their hospitality.

\subsubsection*{Author contribution statement}

\small
Both authors contributed equally to the present work,
were equally involved in the preparation of the manuscript,
and have read and approved the final manuscript.

\subsubsection*{Data availability statement}

\small
Data sharing not applicable to this article.

\subsubsection*{Conflict of interest}

\small
The authors declare no conflict of interest.

\appendix

\section{The ferromagnetic chain}
\label{appferro}

This appendix is devoted to the ferromagnetic Ising chain, whose Hamiltonian reads
\beq
\H=-J\sum_i\s_i\s_{i+1}.
\eeq

This model only exhibits long-range ferromagnetic order at zero temperature,
in agreement with general results recalled above (see~\cite{LL,CS}).
At finite inverse temperature $\beta$, the spin correlation function
\beq
\mean{\s_0\s_n}=(\tanh\beta J)^{\abs{n}}
\label{ferrocor}
\eeq
falls off exponentially with the distance $\abs{n}$ between both spins,
defining thus the correlation length
\beq
\xi=-\frac{1}{\ln\tanh\beta J}.
\eeq
At low temperature $(\beta J\gg1)$, this length diverges exponentially fast, as
\beq
\xi\approx\frac{\e^{2\beta J}}{2}.
\eeq

The Ising chain exhibits an interesting low-temperature scaling regime
when the correlation length $\xi$ and the sample length $L$ are both large and proportional to each other.
The main quantity of interest for our purpose is the mean magnetisation,
\beq
m=\frac{1}{L}\sum_{i=1}^L\s_i.
\eeq
The finite-size scaling behaviour of this quantity in the above scaling regime
has been investigated in detail by Antal et~al.~\cite{antal}.
The mean magnetisation has a non-trivial distribution $f_m(m)$,
depending on the scaling variable
\beq
\z=\frac{L}{2\xi},
\eeq
representing the mean number of domain walls in the system,
and on boundary conditions.

The magnetisation distribution has been worked out in~\cite{antal}
for periodic, antiperiodic and free boundary conditions.
Only the last situation is relevant for the present work.
In this case, we have
\beqa
f_m(m)&=&\frac{\e^{-\z}}{2}\left(\delta(m-1)+\delta(m+1)\right)
\nonumber\\
&+&\frac{\z\e^{-\z}}{2}\left(\I_0(\z v)+\frac{\I_1(\z v)}{v}\right),
\label{faim}
\eeqa
where $\I_0$ and $\I_1$ are modified Bessel functions,
with the notation
\beq
v=\sqrt{1-m^2}.
\eeq
The first line of~(\ref{faim}) is the contribution of the two totally magnetised,
i.e., perfectly ferromagnetically ordered, configurations, having no domain walls.
The Laplace transform (or characteristic function) of the above distribution reads
\beq
\hat f_m(s)=\mean{\e^{sm}}=\e^{-\z}\left(\cosh u+\z\,\frac{\sinh u}{u}\right),
\label{flap}
\eeq
with the notation
\beq
u=\sqrt{s^2+\z^2}.
\eeq
The moments of the total magnetisation can be derived by expanding~(\ref{flap}) as a power series in $s$.
We thus obtain
\beqa
\mean{m^2}&=&\frac{2\z-1+\e^{-2\z}}{2\z^2},
\nonumber\\
\mean{m^4}&=&\frac{3(2\z^2-4\z+3-(2\z+3)\e^{-2\z})}{2\z^4},
\eeqa
and so on.
The key quantity used in the body of this paper is the kurtosis of the mean magnetisation,
\beq
B=\frac{\mean{m^4}}{\mean{m^2}^2}.
\label{bdef}
\eeq
This quantity is sometimes referred to as the Binder parameter or Binder ratio~\cite{binder}.
Its expression in the scaling regime reads
\beq
B=\frac{6(2\z^2-4\z+3-(2\z+3)\e^{-2\z})}{(2\z-1+\e^{-2\z})^2}.
\label{bfss}
\eeq
The kurtosis is an increasing function of $\z$,
interpolating between the following limiting regimes:

\begin{itemize}

\item
For $\z\ll1$, i.e., $L\ll\xi$, the system is nearly perfectly magnetised.
The kurtosis departs from unity as
\beq
B=1+\frac{8\z}{15}+\cdots=1+\frac{4L}{15\xi}+\cdots
\eeq

\item
For $\z\gg1$, i.e., $L\gg\xi$, the system is very disordered,
as a typical configuration comprises many domain walls.
The distribution $f_m(m)$ therefore approaches a narrow Gaussian with variance
\beq
\mean{m^2}\approx\frac{1}{\z}\approx\frac{2\xi}{L}.
\eeq
The kurtosis accordingly departs from the Gaussian value 3 as
\beq
B=3-\frac{3}{\z}+\cdots=3-\frac{6\xi}{L}+\cdots
\eeq

\end{itemize}

\section{The spin-glass chain}
\label{appsg}

This appendix is devoted to the Ising spin-glass chain, whose Hamiltonian reads
\beq
\H=-\sum_iJ_i\s_i\s_{i+1}.
\label{hamg}
\eeq
The model studied in Section~\ref{typical} boils down to this model
in the absence of random fields $(w=0)$.

The attractors (one-spin-flip metastable configurations) of the spin-glass chain
have been studied by Derrida and Gardner~\cite{dg} (see also~\cite{em,li}).
In order to investigate this model,
it is advantageous to perform a gauge transformation of the form~(\ref{gauge}),
where the signs $\eps_i=\pm1$ are so far undetermined.
This transformation leaves the form of the Hamiltonian $\H$ unchanged, as we have
\beq
\H=-\sum_i\w J_i\w\s_i\w\s_{i+1}.
\label{hamg2}
\eeq
In the geometry of an open chain,
the signs $\eps_i$ can be chosen so as to have $\w J_i=\abs{J_i}\ge0$.
This can be achieved by setting $\eps_1=+1$, and then recursively
\beq
\eps_{i+1}=\eps_i\sign J_i.
\eeq
With this construction, the Hamiltonian~(\ref{hamg2}) is that of a random ferromagnet.
This mapping demonstrates in particular that the spin-glass chain is not frustrated.
If all exchange couplings are non-zero,
it has exactly two ground-states where all new spins $\w\s_i$ are ferromagnetically ordered.

\begin{figure}[!ht]
\begin{center}
\includegraphics[angle=0,width=0.9\linewidth,clip=true]{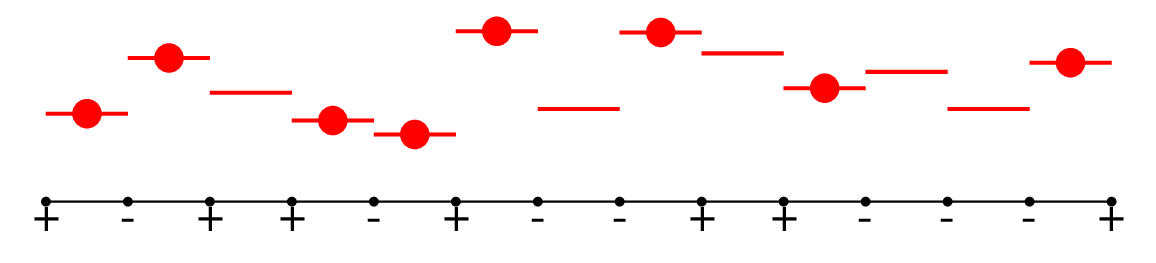}
\vskip 20pt
\includegraphics[angle=0,width=0.9\linewidth,clip=true]{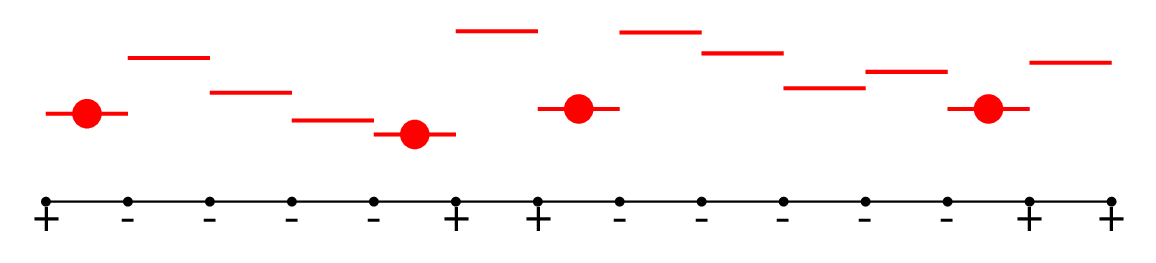}
\caption{\small
An illustration of the descent dynamics~(\ref{dyg}) on a sample of length $L=14$.
Red bars: landscape of exchange couplings.
Red bullets: domain walls, i.e., unsatisfied bonds.
Upper panel: a random configuration of spins.
Lower panel: one of the possible attractors.}
\label{glass}
\end{center}
\end{figure}

In terms of the spins $\w\s_i$,
the zero-temperature dynamical rule~(\ref{dy}) reads
\beq
\w\s_i\to
\left\{
\begin{matrix}
\hfill\w\s_{i-1}\quad &\hbox{if }\w J_{i-1}>\w J_i,\hfill\cr
\hfill\w\s_{i+1}\quad &\hbox{if }\w J_i>\w J_{i-1}.\hfill
\end{matrix}
\label{dyg}
\right.
\eeq
In other words, $\w\s_i$ aligns with the neighbouring spin to which it is the more strongly coupled.
Figure~\ref{glass} illustrates this dynamics on a sample of length $L=14$.
The landscape of exchange couplings $\w J_i$ is shown by red bars.
Red bullets denote domain walls, i.e., unsatisfied bonds, such that $\w\s_i\w\s_{i+1}=-1$.
The upper panel depicts a random configuration of spins,
whereas the lower one depicts
one of the attractors that can be reached from the latter configuration by the dynamics~(\ref{dyg}).
This attractor is henceforth denoted by~$W$.

Under the descent dynamics~(\ref{dyg}),
bullets (representing domain walls) move downhill to weaker neighbouring bonds,
and annihilate pairwise when they meet.
The attractors of the dynamics are therefore configurations where all bullets occupy weak bonds,
i.e., local minima of the landscape of exchange couplings.
For a chain of length $L$ having $M$ weak bonds,
each of these bonds may be either occupied by a bullet or not.
There are therefore
\beq
\N=2^M
\eeq
attractors.
Now, bond number $i$ is weak if $\w J_i$ is the smaller of the three numbers $\w J_{i-1}$, $\w J_i$
and
$\w J_{i+1}$.
Hence, for independent random exchange couplings, the density of weak bonds reads
\beq
\rho=\frac13.
\label{rho}
\eeq
We have therefore $\mean{M}\approx\rho L$, so that
\beq
\mean{\ln\N}\approx S_0L,
\label{logave}
\eeq
with
\beq
S_0=\frac{\ln 2}{3}\approx0.231049.
\label{szero}
\eeq
The full distribution of the number $M$ of weak bonds on large samples
is asymptotically characterised by a large-deviation formula of the form
\beq
\mean{\e^{sM}}\sim\e^{\lambda(s)L},
\label{ldf}
\eeq
with rate function
\beq
\lambda(s)=\ln\frac{\sqrt{\e^s-1}}{\arctan\sqrt{\e^s-1}}.
\eeq
All the above results were first derived in~\cite{dg}.
As a consequence of~(\ref{ldf}),
all cumulants of the number of minima grow linearly with $L$, according to
\beq
\mean{M^k}_c\approx c_kL,
\eeq
with
\beq
\lambda(s)=\sum_{k\ge1}\frac{c_ks^k}{k!},
\eeq
so that
\beq
c_1=\frac{1}{3},\quad
c_2=\frac{2}{45},\quad
c_3=-\frac{2}{945},
\eeq
and so on.
In particular $c_1=\lambda'(s=0)=\rho=1/3$, in agreement with~(\ref{rho}).

As a consequence of the above, the typical number of attractors grows as
\beq
\N_\typ=\e^{\mean{M}\ln 2}\sim\e^{(\ln 2)\lambda'(s=0)L}\sim(2^{1/3})^L
\eeq
(see~(\ref{logave}),~(\ref{szero})),
whereas the mean number of attractors grows as
\beq
\mean{\N}=\mean{\e^{M\ln 2}}\sim\e^{\lambda(s=\ln 2)L}\sim(4/\pi)^L.
\eeq
The difference between the growth rates $2^{1/3}\approx1.259921$ and $4/\pi\approx1.273239$ is significant,
although numerically small in the present case.

\begin{figure}[!ht]
\begin{center}
\includegraphics[angle=0,width=0.9\linewidth,clip=true]{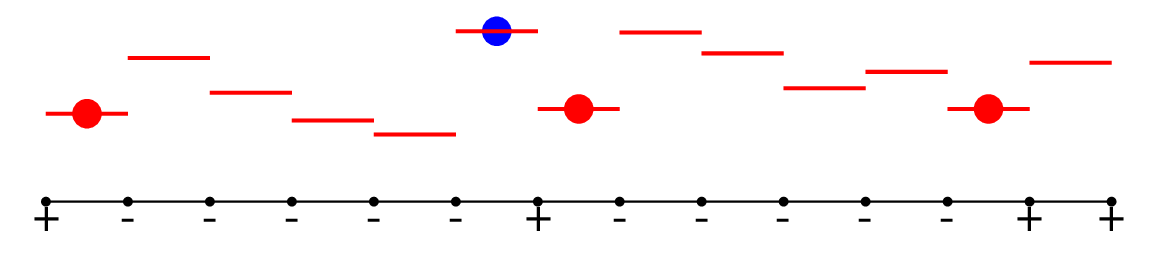}
\caption{\small
An example of a single mishearing that might not be healed by the decryption algorithm.
The spin configuration is obtained by flipping $\w\s_6$ in the attractor $W$.
This flip has suppressed one of the red bullets and excited the blue one.}
\label{gflip}
\end{center}
\end{figure}

The above characterisation of attractors can be extended in order to predict the exact values
of the amplitudes $a_0$ and $b_0$ entering~(\ref{npsca}) in the absence of random fields $(w=0)$.
The estimates~(\ref{npsca}) hold in the regime where the density $p$ of initial mishearings is small.
It is therefore sufficient to consider the effect of introducing a single mishearing,
i.e., of flipping a single spin at a uniform random location into an attractor~$W$.
Such a spin flip generates two neighbouring bullets.
The decryption stage will be able to heal the mishearing if
these bullets annihilate at some later time.
Conversely, the mishearing will not be healed, and $X$ will be different from $W$,
whenever the bullets have a different fate.
This may occur if the flipped spin is adjacent to a strong bond,
i.e., a local maximum of the landscape of exchange couplings.
The configuration shown in Figure~\ref{gflip} provides such an instance.
It is obtained by flipping the spin $\w\s_6$ in the attractor~$W$
shown in the lower panel of Figure~\ref{glass}.
The flipped spin is the left end of a strong bond.
As a result, one of the red bullets has disappeared,
whereas another bullet (shown in blue) occupies the strong bond between $\w\s_6$ and $\w\s_7$.

The fate of the blue bullet determines whether the mishearing will be healed or not.
If $\w\s_6$ is updated before $\w\s_7$, the blue bullet moves left,
the attractor $W$ of the lower panel of Figure~\ref{glass} is recovered,
and the mishearing is healed.
If $\w\s_7$ is updated before $\w\s_6$, the blue bullet moves right,
annihilates with the first red bullet to its right,
and an attractor $X$ different from~$W$ is reached.
In this situation, $X$ can be derived from~$W$ by flipping all spins between two consecutive weak bonds,
i.e., two consecutive minima of the exchange coupling landscape.
The random number $N$ of residual mishearings (see~(\ref{ndef}))
therefore identifies with the distance between consecutive minima,
whose distribution is known exactly~\cite[Eq.~(23)]{dg}:
\beq
\prob{N=k}=\frac{3(k-1)(k+2)2^k}{(k+3)!}\quad(k\ge2).
\label{pdist}
\eeq
We have in particular
\beq
\mean{N}=\frac{1}{\rho}=3,\quad
\mean{N^2}=3\e^2-12\approx10.167168.
\label{nmoms}
\eeq

The statistics of the above histories are as follows.
First, weak and strong bonds interlace along the chain and have the same density $\rho$ (see~(\ref{rho})).
For ordered spin updates,
$\w\s_6$ is always updated before $\w\s_7$.
Mishearings sitting to the left of strong bonds are thus always healed,
whereas those sitting to the right of strong bonds are never healed.
For random spin updates, $\w\s_6$ is updated before $\w\s_7$ with probability $1/2$.
Both kinds of mishearings are therefore healed with probability $1/2$.
Ordered and random updates therefore have exactly the same efficiency in the present situation.
Finally, whenever the mishearing is not healed, the distance between $W$ and $X$,
equal to the number~$N$ of residual mishearings after the decryption stage has been performed,
has the distribution~(\ref{pdist}).
This non-frustrated model is the only situation where the distribution of~$N$ is known.
We thus arrive at the following predictions
\beq
b_0=\rho=\frac13,\quad
a_0=\rho\mean{N}=1,
\label{abzero}
\eeq
that are common to both updating schemes considered in this work.

\section{The $g\to\infty$ limit}
\label{appg}

This appendix is devoted to the $g\to\infty$ limit of the model describing long words,
studied in Section~\ref{long}.
In this limit, the total fields $\eta_i$ defined in~(\ref{etalong})
are dominated by the variable-range symmetry-breaking component, which is proportional to $g$, namely
\beq
\eta_i\approx g\sum_{j=i+1}^Lx^{j-i-1}\s_j,
\label{etagene}
\eeq
and in particular
\beq
\eta_{L-1}\approx g\s_L.
\label{eta1}
\eeq
The rightmost spin $(i=L)$ constitutes an exception to the rule~(\ref{etagene}), namely
\beq
\eta_L=J\s_{L-1}+h,
\label{eta0}
\eeq
where we have introduced the shorthand notations
\beq
J=J_{L-1},\quad h=h_L.
\eeq
The fixed points of the dynamics, obeying~(\ref{fp}), can be determined as follows.
For such a fixed point,~(\ref{eta1}) implies $\s_{L-1}=\s_L$, so that~(\ref{eta0}) yields
\beq
\s_L=\sign(J\s_L+h).
\eeq
The solution $\s_L=+1$ holds for $J+h>0$.
We have then $\s_i=+1$ for all $i$.
Similarly, the solution $\s_L=-1$ holds for $J-h>0$.
We have then $\s_i=-1$ for all $i$.

\begin{figure}[!ht]
\begin{center}
\includegraphics[angle=0,width=0.9\linewidth,clip=true]{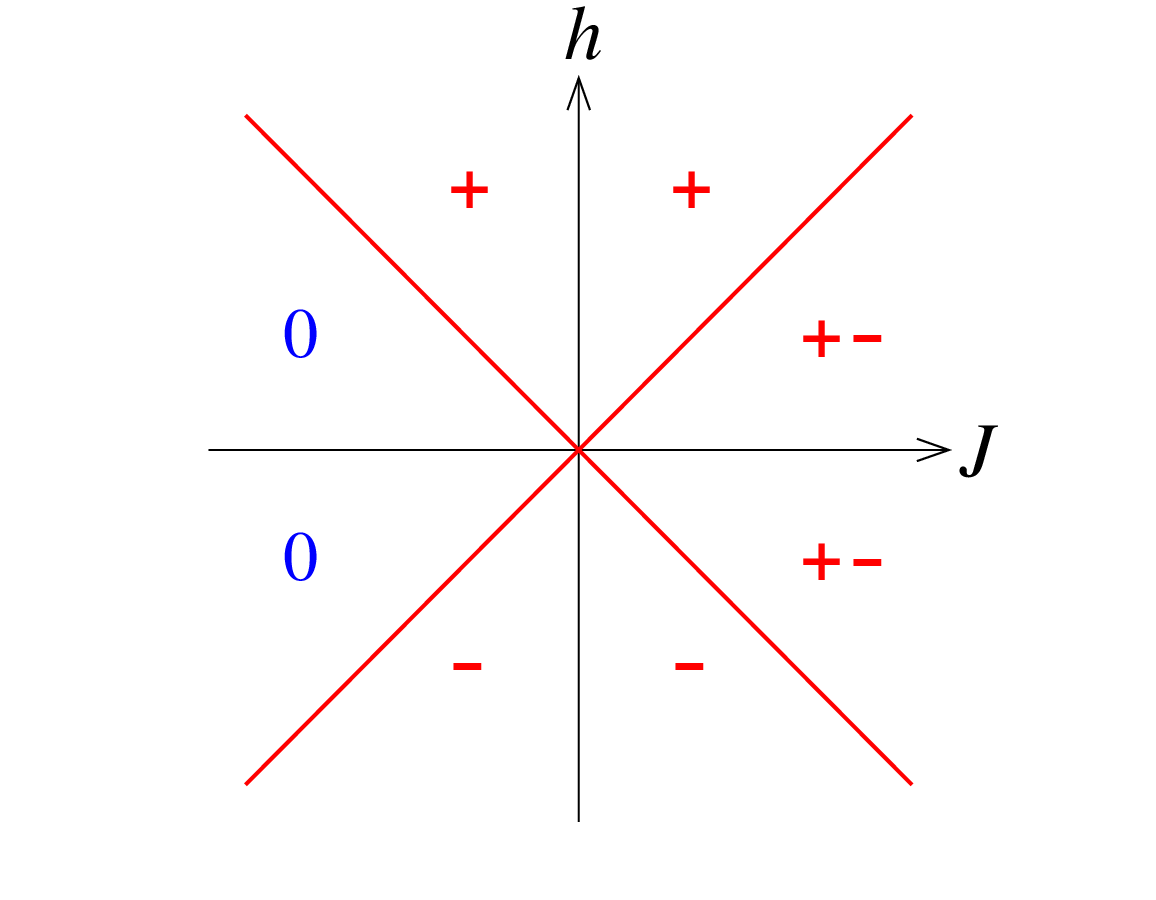}
\caption{\small
Number and structure of the fixed points of the dynamics in the $J$--$h$ plane.
Symbols $+$ and $-$: regions with on single fixed point, $\s_i=+1$ or $\s_i=-1$.
Symbols $+-$: regions where both fixed points coexist.
Symbols 0: regions with no fixed point.}
\label{jhplot}
\end{center}
\end{figure}

Depending on the values of $J$ and $h$,
there are thus 0, 1 or 2 fixed points in the $g\to\infty$ limit (see Figure~\ref{jhplot}).
These fixed points are ferromagnetically ordered.
It is important to note that this limiting situation is therefore, in all respects, far from the
regime where fixed points can be identified with realistic words.

The probability of having no fixed points reads
\beq
P_0=\prob{J<0\hbox{ and }\abs{h}<-J}.
\eeq
The probabilities $P_1$ and $P_2$ of there being one or two fixed points then read, by symmetry:
\beq
P_1=1-2P_0,\qquad P_2=P_0.
\eeq
With the symmetric distributions~(\ref{rhos}) of $J$ and~$h$, we have
\beq
P_0=\left\{
\begin{matrix}
\frac{2-w}{4}\hfill\quad & (w\le1),\hfill\cr
\frac{1}{4w}\hfill\quad & (w\ge1).\hfill
\end{matrix}
\right.
\eeq
For $w=1$, the value usually used in this work,
the distribution of the number of fixed points simplifies to
\beq
P_0=\frac14,\quad P_1=\frac12,\quad P_2=\frac14.
\eeq

\section{The $\ell\to0$ limit}
\label{appell}

This appendix is devoted to the $\ell\to0$ limit of the model describing long words, studied in Section~\ref{long}.
In the latter limit, the total fields $\eta_i$ are given by~(\ref{etazero}), i.e.,
\beq
\eta_i=J_{i-1}\s_{i-1}+(J_i+g)\s_{i+1}+h_i,
\label{etaxi0}
\eeq
again with the exception~(\ref{eta0}) for $i=L$.

The fixed points of the dynamics~(\ref{dylong}) remain those shown in Figure~\ref{jhplot}
as long as the term proportional to $g$ is dominant in~(\ref{etaxi0}).
With the symmetric distributions~(\ref{rhos}) of $J$ and~$h$, this condition reads $g>g_0$, with
\beq
g_0=w+2.
\eeq
For $w=1$, this reads
\beq
g_0=3.
\label{g0}
\eeq

For values of $g$ below $g_0$, the attractors become progressively more numerous,
and they exhibit short-range ferromagnetic order,
characterised by some correlation length $\xi$
which is expected to diverge as $g\to g_0$.
This behaviour can be analysed as follows.
Consider a fixed point of the dynamics, obeying $\eta_i\s_i>0$ for all $i$ (see~(\ref{fp})), i.e.,
\beq
J_{i-1}\s_{i-1}\s_i+(J_i+g)\s_i\s_{i+1}+h_i\s_i>0.
\label{sineqs}
\eeq
For a fixed position $i<L$ along the chain,
and with the symmetric distributions~(\ref{rhos}) of $J_i$ and~$h_i$,~(\ref{sineqs}) yields
\beq
\mean{\s_i\s_{i+1}}=1-2\,\prob{y>g},
\label{aimprob}
\eeq
where the random variable
\beq
y=J_{i-1}+J_i+h_i
\eeq
has an even distribution $f_y(y)$ on $-g_0\le y\le g_0$.

From now on, we focus our attention onto $w=1$, the value usually used in this work.
We then have
\beq
f_y(y)=\left\{
\begin{matrix}
\frac{1}{16}(3+y)^2\hfill\quad & (-3\le y\le -1),\hfill\cr
\frac{1}{8}(3-y^2)\hfill\quad & (-1\le y\le 1),\hfill\cr
\frac{1}{16}(3-y)^2\hfill\quad & (1\le y\le 3),\hfill\cr
\end{matrix}
\right.
\eeq
so that~(\ref{aimprob}) reads
\beq
\mean{\s_i\s_{i+1}}=1-\frac{(3-g)^3}{24}\qquad(1\le g\le 3).
\eeq
Identifying this result with the expression~(\ref{ferrocor})
for the correlation function of the ferromagnetic chain at equilibrium yields
the following approximate expression for the correlation length:
\beq
\xi\approx-\frac{1}{\ln\left(1-\frac{(3-g)^3}{24}\right)}.
\label{xieq}
\eeq
Finally, setting $\xi=100$ in the above yields
\beq
g_c\approx2.379587.
\label{gcz}
\eeq

\bibliography{speech.bib}

\end{document}